%% file: main.tex
\documentclass[sigconf, usenames, dvipsnames]{acmart}

\usepackage{tabularx}
\usepackage{xcolor}
\usepackage{tasks}
\usepackage{enumitem}
\usepackage{wrapfig}
\usepackage{graphicx}
\usepackage{subcaption}
\usepackage{booktabs}
\usepackage{cancel}
\hypersetup{hypertexnames=false}
\newcommand{\revision}[1]{{\color{black} #1}}
\newcommand{\remove}[1]{}

\copyrightyear{2025}
\acmYear{2025}
\setcopyright{acmlicensed}\acmConference[CHI '25]{CHI Conference on Human
Factors in Computing Systems}{April 26-May 1, 2025}{Yokohama, Japan}
\acmBooktitle{CHI Conference on Human Factors in Computing Systems (CHI
'25), April 26-May 1, 2025, Yokohama, Japan}
\acmDOI{10.1145/3706598.3713293}
\acmISBN{979-8-4007-1394-1/25/04}

\AtBeginDocument{
  \providecommand\BibTeX{{%
    \normalfont B\kern-0.5em{\scshape i\kern-0.25em b}\kern-0.8em\TeX}}}

\begin{document}
\title{From Following to Understanding: Investigating the Role of Reflective Prompts in AR-Guided Tasks to Promote Task Understanding}

\author{Nandi Zhang}
\affiliation{%
\institution{University of Calgary}
\city{Calgary}
\country{Canada}}
\email{nandi.zhang@ucalgary.ca}

\author{Yukang Yan}
\orcid{0000-0001-7515-3755}
\affiliation{%
\institution{University of Rochester}
\city{Rochester}
\country{USA}}
\email{yukang.yan@rochester.edu}

\author{Ryo Suzuki}
\orcid{0000-0003-3294-9555} 
\affiliation{%
\institution{University of Colorado Boulder}
\city{Boulder}
\country{USA}}
\email{ryo.suzuki@colorado.edu}

\renewcommand{\shortauthors}{Zhang, Yan, and Suzuki}

\input{0-abstract}

\begin{teaserfigure}
\includegraphics[width=\textwidth]{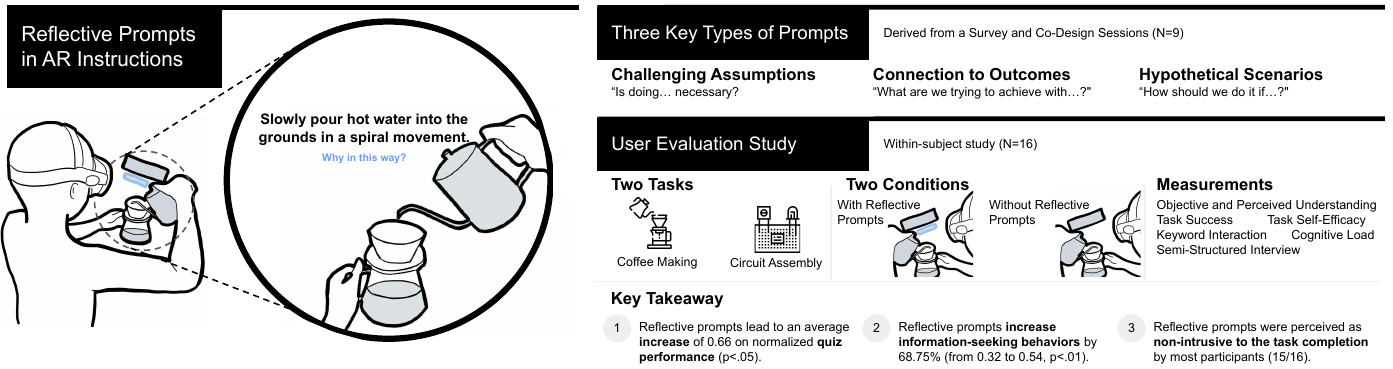}
\caption{Left: Example of an AR instruction (in black texts) with a reflective prompt (in blue texts) for a coffee-making task. Right: Study design framework showing (a) Three key types of reflective prompts derived from the formative study, (b) User evaluation study design with two tasks and two conditions, and (c) Key findings demonstrating the effectiveness of reflective prompts in AR instructions.}
\label{fig:teaser}
\end{teaserfigure}

\maketitle

\input{1-introduction}
\input{2-related-work}
\input{3-formative-study}
\input{4-evaluation-study}
\input{5-design-guidelines}

\input{6-discussion}
\input{7-conclusion}

\input{acknowledgements}

\balance
\bibliographystyle{ACM-Reference-Format}
\bibliography{references}

\end{document}

%% file: 0-abstract.tex
\begin{abstract}
Augmented Reality (AR) is a promising medium for guiding users through tasks, yet its impact on fostering deeper task understanding remains underexplored. This paper investigates the impact of \textit{reflective prompts}---strategic questions that encourage users to \textit{challenge assumptions}, \textit{connect actions to outcomes}, and consider \textit{hypothetical scenarios}---on task comprehension and performance. We conducted a two-phase study: a formative survey and co-design sessions (N=9) to develop reflective prompts, followed by a within-subject evaluation (N=16) comparing AR instructions with and without these prompts in coffee-making and circuit assembly tasks. Our results show that reflective prompts significantly improved objective task understanding and resulted in more proactive information acquisition behaviors during task completion. These findings highlight the potential of incorporating reflective elements into AR instructions to foster deeper engagement and learning. Based on data from both studies, we synthesized design guidelines for integrating reflective elements into AR systems to enhance user understanding without compromising task performance.
\end{abstract}

\ccsdesc[500]{Human-centered computing~Mixed / augmented reality}

\keywords{\revision{Augmented Reality; Task Guidance; Instruction Following; Reflective Prompts}}

%% file: 1-introduction.tex
\section{Introduction}
In recent years, Augmented Reality (AR) technology has emerged as a powerful tool to enhance task guidance. By overlaying digital content onto the physical world, AR provides real-time, contextual guidance for various tasks, from complex industrial procedures to everyday activities. AR instructions have been successfully implemented in various scenarios, including cooking \cite{castelo2023argus, zhai2020interactive, wu2024artist}, assembly tasks \cite{yang2019influences, funk2016interactive, biocca2007attention, hou2013using, deshpande2018effects}, or even complex manual tasks \cite{dhiman2024does}. 

While AR-based instructions offer many advantages, they could potentially lead users to passively follow the steps without deep reasoning~\cite{radu2019}, a limitation compared to human experts or well-developed educational manuals. In traditional settings, human experts provide step-by-step instructions and foster deeper understanding through \textit{reflective prompts}---questions or statements designed to provoke thoughtful reflection and deeper understanding \cite{davis2000scaffolding, guo2022should}. For instance, when assembling a piece of furniture, an expert may ask, ``Why do you think it's important to tighten the screws gradually and evenly?'', which may encourage the learner to think about the importance of balance and structure in the assembly process. These reflective prompts engage users more deeply, helping them develop task understanding and eventually transferable skills—an element often missing in AR-guided task completion.

In this paper, we explore how reflective prompts can be integrated into AR instructional systems. Specifically, we aim to answer the following research question: \textbf{RQ: Can reflective prompts be effectively incorporated into AR task guidance to enhance both task completion and deeper understanding?} To address this, we break the question down into two sub-questions: 1) \textbf{RQ1: What types of reflective prompts (e.g., critical reflection vs. perspective shifting) are effective and appropriate in an AR instruction-following context?} and 2) \textbf{RQ2: What are the impacts AR-based reflective prompts have on task performance and user understanding compared to AR instructions without such elements?}

To answer RQ1, we first conducted a formative study where we extracted five categories of reflective prompts from literature in learning and education fields and implemented them in a prototype system on an AR head-worn display (Apple Vision Pro\footnote{https://www.apple.com/apple-vision-pro/}). We then engaged nine participants in a co-design process, having them perform two tasks---making pour-over coffee and assembling circuits on a breadboard---using our system. Through iterative design sessions, participants helped us refine the choice of reflective prompts. Ultimately we identified three types of reflective prompts that are most effective and preferable for the AR instruction-following context---\textit{Challenging Assumptions, Connections to Outcomes, Hypothetical Scenarios}. 

To address RQ2, we conducted an evaluation study with 16 participants to assess the impact of these refined prompts on task performance, user understanding, and their willingness to seek additional task-related information.
Results show that AR instructions with reflective prompts significantly improved objective understanding which led to significantly higher scores (p<.05) in quizzes on tasks. \revision{Reflective prompts also fostered \textbf{epistemic curiosity}—a cognitive trait that motivates learning and enhances knowledge retention~\cite{kang2009wick, murayama2019process}. This was evidenced by participants' voluntary information-seeking behavior~\cite{kidd2015psychology, berlyne1954theory, wilson2024curiosity}, with those in the reflective prompt condition proactively seeking significantly more additional task-related information (p<.05).} \remove{Reflective prompts also encouraged participants to seek additional task-related information and resulted in significantly more proactive information acquisition behaviors (p<.05). }The insights gained from both the formative and evaluation studies informed the development of design guidelines that enhance both immediate task performance and understanding of AR-guided tasks.

In this paper, we contribute to the understanding of how reflective elements can be integrated into AR task guidance through the following:

\begin{enumerate}
    \item A formative study (N=9) that identified gaps in instruction-following within AR contexts and adapted reflection types from existing literature into effective reflective prompts.
    \item An evaluation study (N=16) that assessed the impact of incorporating reflective prompts into AR instructions on task performance, user understanding, and information-seeking behavior.
    \item \remove{Design guidelines that provide implications}\revision{Design implications that include design guidelines} for the design of future \revision{reflective} AR instructional systems \remove{that balance immediate task assistance with fostering deeper task understanding}\revision{and design considerations for applying them to broader task scenarios}.
\end{enumerate}

%% file: 2-related-work.tex
\section{Related Work}
\subsection{Task Guidance in AR}
AR has been widely explored as a task guidance medium in domains, such as cooking \cite{castelo2023argus, zhai2020interactive, wu2024artist}, object assembly \cite{yang2019influences, funk2016interactive, biocca2007attention, hou2013using, lampen2019combining, deshpande2018effects, alves2019comparing}, surgery \cite{meola2017augmented, andersen2018augmented}, military \cite{boyce2019impact, laviola2015using, ejder2012augmented}, maintenance \cite{fiorentino2014augmented, obermair2020maintenance, wang2022usability}. 
By overlaying instructions directly onto the workspace, AR reduces attention shifts between instructional materials and the task \revision{\cite{polvi2017handheld, buchner2022impact}}, potentially lowering the cognitive load and motivating learning \revision{\cite{kuccuk2016learning, lai2019augmented, funk2016interactive, bellucci2018investigating, chiang2014augmented, mao2017impact}}, and improving task performance \revision{\cite{kuccuk2016learning, yang2019influences, wenk2019reaching, lampen2019combining}} \remove{\cite{buchner2022impact, chiang2014augmented, kuccuk2016learning, lai2019augmented, mao2017impact, polvi2017handheld, yang2019influences, funk2016interactive, wenk2019reaching, lampen2019combining, bellucci2018investigating}}. 
Studies have shown that AR can lead to faster task completion, reduced error rates, and increased user satisfaction compared to traditional methods like paper manuals or screen-based instructions \cite{wang2021role, buchner2022impact, wang2022usability, lampen2019combining}.

However, some research has reported contrary findings, suggesting that AR may increase cognitive load or lead to poorer performance in certain contexts \cite{he2019ar, kawai2010ergonomic, sedighi2018information, deshpande2018effects, pu2018development}.  Additionally, while AR simplifies task performance through step-by-step guidance, it may reduce users' ability to detect prior mistakes or adapt to real-world variables, leading to lower knowledge retention and over-reliance on the system \cite{tang2003comparative, rehman2016augmented, warden2023fast}.

While prior studies primarily focus on comparing AR to other instructional media in terms of task performance and cognitive load, our research explores how features within AR---such as reflective prompts---can address these limitations. By encouraging users to reflect and engage with the underlying principles of a task, we seek to mitigate issues like blind adherence to instructions and poor knowledge retention, ultimately enhancing both task performance and deeper understanding.

\subsection{Reflective Prompts in Educational Contexts}
The role of reflection in fostering deeper learning and understanding has been extensively emphasized in psychology and educational theory. Reflection enables learners to make sense of their experiences by synthesizing specific actions into broader principles and abstract concepts \cite{kolb2014experiential, daudelin1996learning, wain2017learning}. Through reflective thinking, learners can identify patterns in experiences, generalize insights, and apply learned knowledge to new contexts \cite{boyd1983reflective, boud2013reflection}. Moreover, reflection encourages learners to critically assess their decisions, actions, and assumptions, prompting them to actively question their values and thought processes \cite{chang2019reflection, costa2008learning}. This form of metacognitive engagement often leads to more insightful learning outcomes and a greater ability to transfer knowledge across different domains \cite{kolb2014experiential}.

In educational settings, instructors frequently use reflective prompts and feedback mechanisms to guide students through this reflective process \cite{stark2009effects, davis2000scaffolding, menekse2022different, lai2006effect}. Such prompts are designed to encourage students to pause, think critically about their actions, and make connections between theoretical concepts and practical experiences \cite{roskos2001reflection, bannert2006effects, davis2000scaffolding, krause2010reflection}. For instance, prompts may ask students to review what they have done, explain their reasoning, justify their decisions, or explore alternative approaches to a problem \cite{roskos2001reflection, davis2000scaffolding, choi2023benefit}. These prompts often serve as scaffolding to help students navigate complex tasks, ultimately fostering a deeper understanding of the subject matter. By incorporating structured reflection into the learning process, educators aim to transform passive learning into an active, inquiry-driven experience \cite{lai2006effect, powell2005conceptualising, menekse2022different}.

\revision{Studies have also explored reflective components in AR-based learning. Chen et al. found students showed stronger flow states in digital games for science learning versus AR games with reflection prompts \cite{chen2020impacts}, while Lin et al. demonstrated that teacher-guided reflection in AR-based scientific learning improved learning outcomes \cite{lin2023effects}.} However, applying reflective prompts in \remove{instructional}\revision{instruction-following} contexts---such as AR-based task guidance---presents unique challenges. Unlike\remove{traditional} educational \remove{environments}\revision{settings}, where learners are typically students motivated to engage in reflective thinking, users in task-focused contexts may have different goals, preferences, and mindsets. Users of AR guidance systems might primarily focus on completing tasks efficiently rather than on skill development or knowledge retention. As a result, reflective prompts in such settings need to be carefully designed to align with users' immediate goals without disrupting their task flow. Prompts that are too intrusive or cognitively demanding may frustrate users.

\vspace{-5pt}
\subsection{Learning by Doing}
Reflection plays a crucial role in transforming experience into knowledge or skill \cite{daudelin1996learning, boyd1983reflective, boud2013reflection}.  \revision{As Dewey stated, ``We do not learn from experience. We learn from reflecting on experience''~\cite{dewey2022we}, emphasizing that thoughtful reflection is essential for learning. Similarly, Kolb’s Experiential Learning Theory (ELT) posits that knowledge is generated through the transformation of experience \cite{kolb2014experiential}.}\remove{According to Kolb’s Experiential Learning Theory (ELT), knowledge is generated through the transformation of experience \cite{kolb2014experiential}.} This model emphasizes that learning is not simply about completing tasks or accumulating experiences. Instead, it involves an iterative process of engaging with experiences, reflecting on them, conceptualizing insights, and applying those insights to new situations. Without reflection, individuals may successfully complete tasks but fail to internalize the underlying principles and processes \cite{boud2013reflection}.

As \remove{\cite{boud2013reflection}}\revision{Boud et al.} argues \revision{\cite{boud2013reflection}}, reflection is not a passive activity, where one merely recalls past experiences. Rather, it is an active, deliberate process that involves probing experiences, making sense of them, and extracting meaningful learning outcomes. In addition to post-experience reflection, \remove{\cite{schon2017reflective}}\revision{Sch\"{o}n and Donald A} introduced the concept of reflection-in-action \revision{\cite{schon2017reflective}}, where learners reflect and adapt their actions in real time, making necessary adjustments as they proceed. In this paper, we explore how these reflective practices can be integrated into the task-completion process with AR guidance.

%% file: 3-formative-study.tex
\section{Formative Study}
To inform the design of reflective prompts, we first conducted a formative survey on reflection practice in the learning and education fields. Though not specific to AR-guided tasks, this survey offered valuable insights potentially adaptable to AR environments. \revision{We examined 88 papers identified through Google Scholar and ERIC using search terms including ``reflection,'' ``reflective,'' ``learning,'' ``pedagogical,'' ``questioning,'' ``prompts,'' ``guidance,'' and ``instructions'', and their meaningful combinations. Through citation tracking, we identified eight highly influential papers, each with more than 5,000 citations, that could potentially inform reflection design in AR-guided tasks.} \remove{From the literature, we identified}\revision{From the eight papers, we extracted} five main types with eleven subtypes of reflection processes, which served as a starting point for participants to brainstorm and iterate on the design of reflective prompts in our context (as shown in Fig. \ref{fig:formative}).
Then we incorporated \remove{the identified five main types with eleven subtypes of reflection}\revision{all the reflection types} as prompts in an AR instructional system prototype.
Through a co-design session with nine participants, we let them perform two tasks following the instructions from the AR system, which included different types of reflective prompts.
By analyzing the preference of participants for these prompts, we aimed to address the \textbf{[RQ1]}: What types of reflective prompts are effective and appropriate in an AR instruction-following context?

\subsection{Survey of Reflective Prompts} \label{Survey}
We conducted a formative survey of literature on reflection in educational contexts, focusing on how \remove{students}\revision{learners} engage in reflective practices and how educators support them. Though not specific to AR-guided tasks, this survey offered valuable insights potentially adaptable to AR environments. From the literature, we identified five main types with eleven subtypes of reflection processes, which served as a starting point for participants to brainstorm and iterate on the design of reflective prompts in our context (as shown in Fig. \ref{fig:formative}).

\subsubsection*{\textbf{[R1] Critical Reflection:}} Critical reflection involves challenging assumptions and engaging in deeper logical reasoning about tasks and processes. Grounded in Mezirow’s Transformative Learning Theory \cite{mezirow1990fostering} and King and Kitchener’s Reflective Judgment Model \cite{king1994developing}, these prompts encourage users to reconsider their actions and underlying beliefs.

\subsubsection*{\textbf{[R2] Reflective Pauses:}} Inspired by Schön’s Reflection-on-action concept \cite{schon2017reflective}, reflective pauses introduce intentional breaks in instruction for users to process and reflect. These pauses are particularly useful in procedural tasks, allowing users to step back and assess their actions \cite{schon2017reflective}.

\subsubsection*{\textbf{[R3] Metacognitive Awareness:}} Based on Zimmerman’s Self-Regulated Learning Theory \cite{zimmerman2002becoming} and Gibbs’ Reflective Cycle \cite{gibbs1988learning}, these prompts aim to increase users’ awareness of their cognitive processes. By encouraging self-assessment at key points, they enhance task performance and deepen understanding \cite{zimmerman2002becoming}.

\subsubsection*{\textbf{[R4] Perspective Shifting:}} Drawing from Brookfield’s Critical Reflection Model \cite{brookfield2017becoming}, perspective-shifting prompts encourage users to consider hypothetical scenarios and potential context to transfer the skill. This type of reflection can improve problem-solving and broaden understanding in procedural tasks \cite{brookfield2017becoming}.

\subsubsection*{\textbf{[R5] Connection Identification:}} Rooted in Kolb’s Experiential Learning Theory \cite{kolb2014experiential} and Ausubel’s Assimilation Theory \cite{ausubel1968cognitive}, these prompts help users activate prior knowledge, draw connections between tasks and understand how a particular step contributes to the outcome. This reflection supports the integration of new information and enhances knowledge transfer.

\subsection{Co-Design Session}
While Section \ref{Survey} addresses part of RQ1---identifying types of reflective prompts based on the literature---further investigation was needed to explore their effectiveness and appropriateness within an AR instruction-following context. We conducted a co-design session to iterate and refine these prompt categories.

\subsubsection{Participants}
We recruited 9 participants (\revision{F1-9, }5 female, 4 male, age: M=25.6, SD=9 years) from \revision{local universities} \remove{a local community} to participate in a co-design session. Participants were screened to ensure they had no prior experience in the tasks they would be performing. All participants had high-school-level circuit knowledge but no experience with breadboard assembly. Seven of the participants had no prior experience with AR technology.

\subsubsection{Instructional System}
We developed a prototype AR instructional system on the Apple Vision Pro\remove{, a state-of-the-art AR head-worn display}. The system displayed a series of step-by-step instructions as white text overlays within the user's field of view.

The experimenter manually controls when to display and update the instructions using a control interface on a laptop, which communicates with the AR system via a WebSocket connection. Instructions advanced either when the experimenter deemed it appropriate or at the participant’s request. Reflective prompts were toggled by the experimenter.

\begin{figure}
    \centering
    \includegraphics[width=\linewidth]{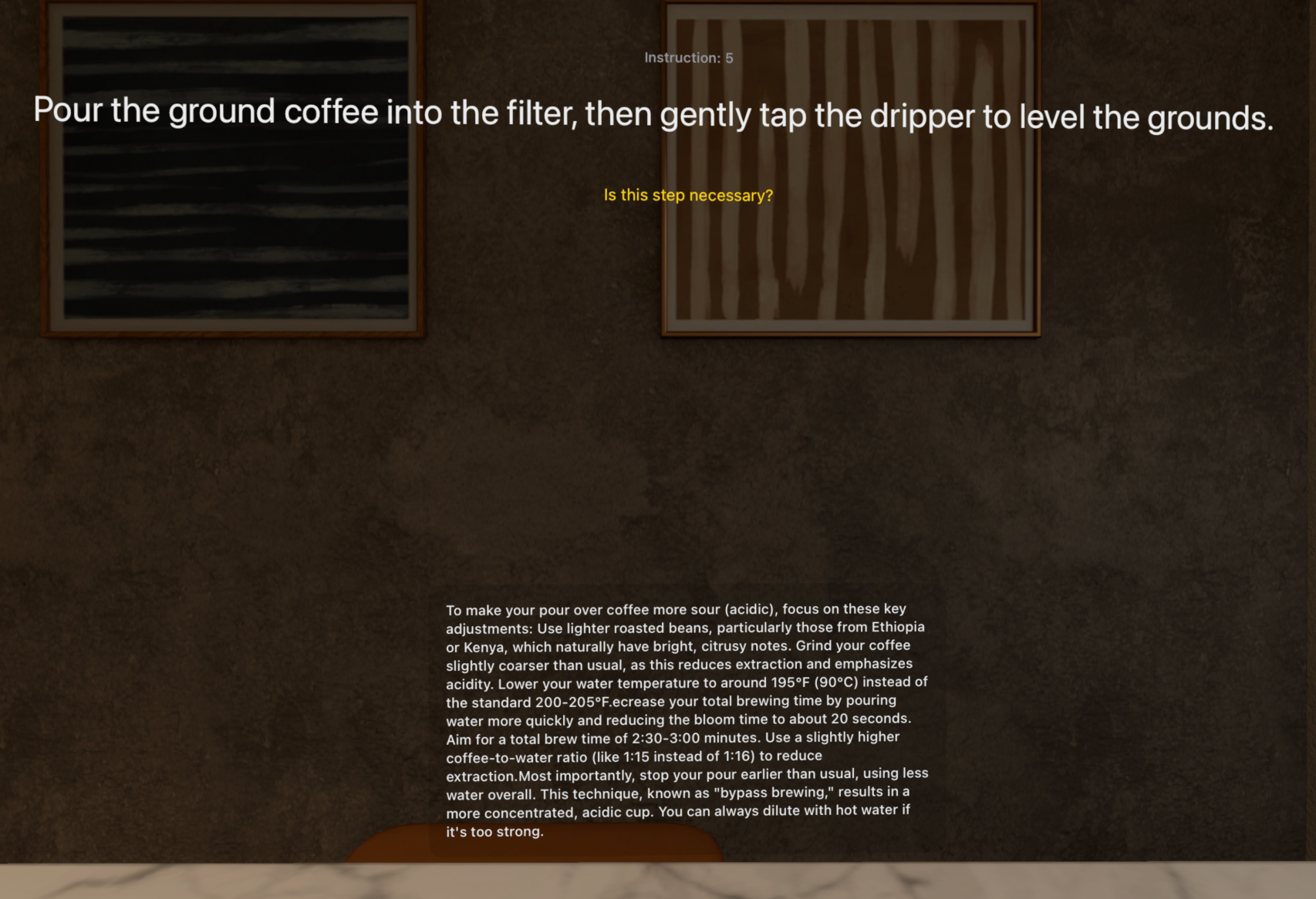}
    \caption{Formative study prototype system layout displaying one step of the Task A instructions (larger white text at the top), reflective prompts (small yellow text underneath) and ChatGPT's response (small white text at the bottom).}
    \label{fig:chatgptscreenshot}
\end{figure}

\revision{We used information-seeking behavior as a measure of epistemic curiosity and learning motivation \cite{kang2009wick, berlyne1954theory, wilson2024curiosity}. While following instructions, users typically seek task-related information through various means---asking others, searching online, or consulting manuals---either to satisfy curiosity or resolve confusion. To capture these natural behaviors in our study environment, we integrated a question-answering system powered by ChatGPT-4o\footnote{https://chat.chatbotapp.ai/?model=gpt-4o} into the AR interface.} Participants could ask task-related questions, which were forwarded to ChatGPT-4o with the responses displayed beneath the instructions (as shown in Fig. \ref{fig:chatgptscreenshot}). This ``wizard-of-oz'' (WoZ) approach simulated natural, seamless information-seeking in AR\remove{. In real-world scenarios, users often hesitate to search for additional information on smartphones due to the high interaction costs and the need to switch between apps. By integrating this functionality directly into the AR environment, we allowed participants to access task-related information on-demand within the system}\revision{, enabling us to measure how reflective prompts influence information-seeking behavior.}

\subsubsection{Tasks}
\begin{figure}[t]
    \centering
    \subfloat[Task A: Making Pour-Over Coffee]{
        \includegraphics[width=0.45\columnwidth]{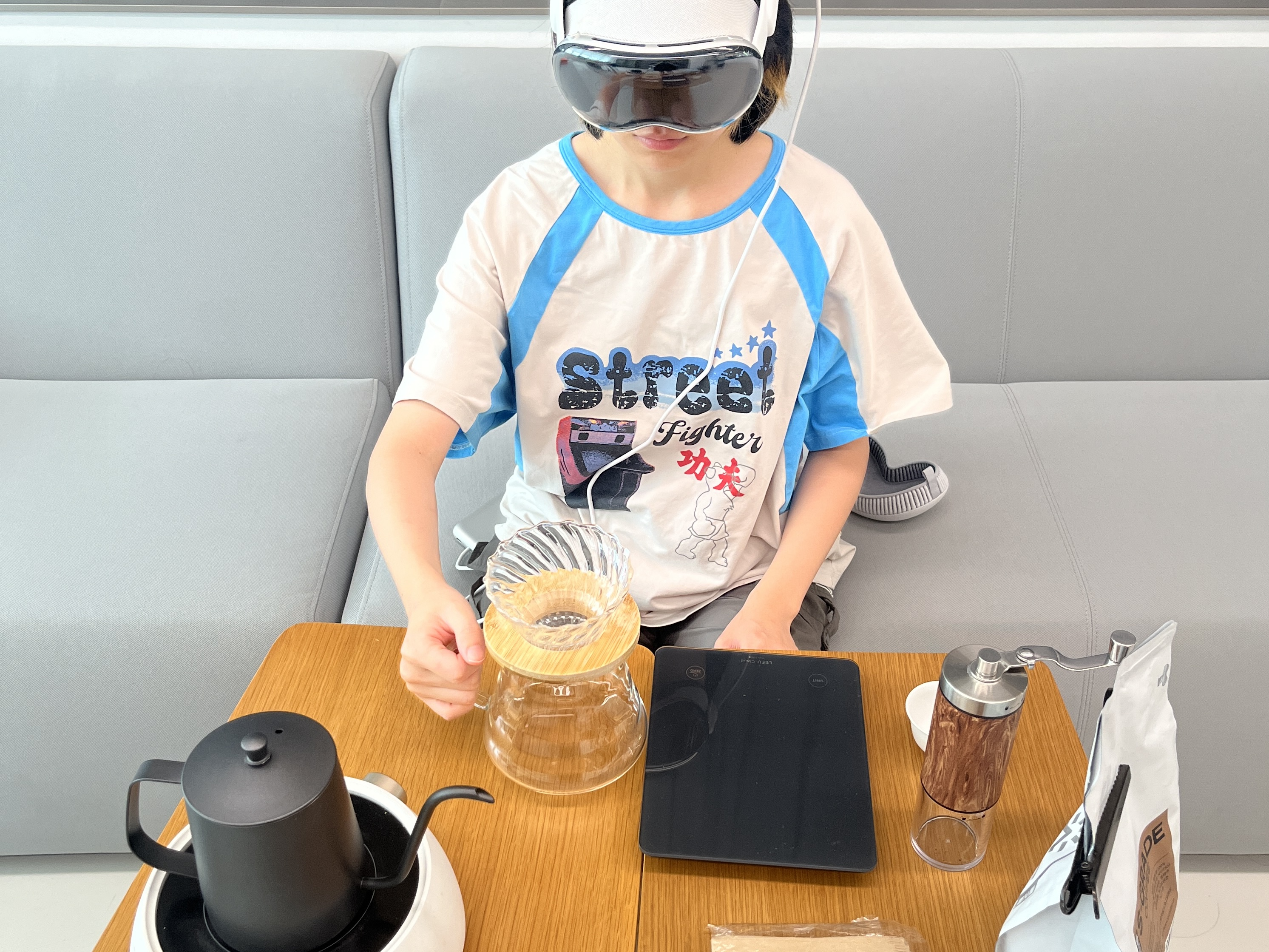}
        \label{fig:coffee}
    }
    \hspace{0.02\columnwidth}
    \subfloat[Task B: Assembling Circuits on a Breadboard]{
        \includegraphics[width=0.45\columnwidth]{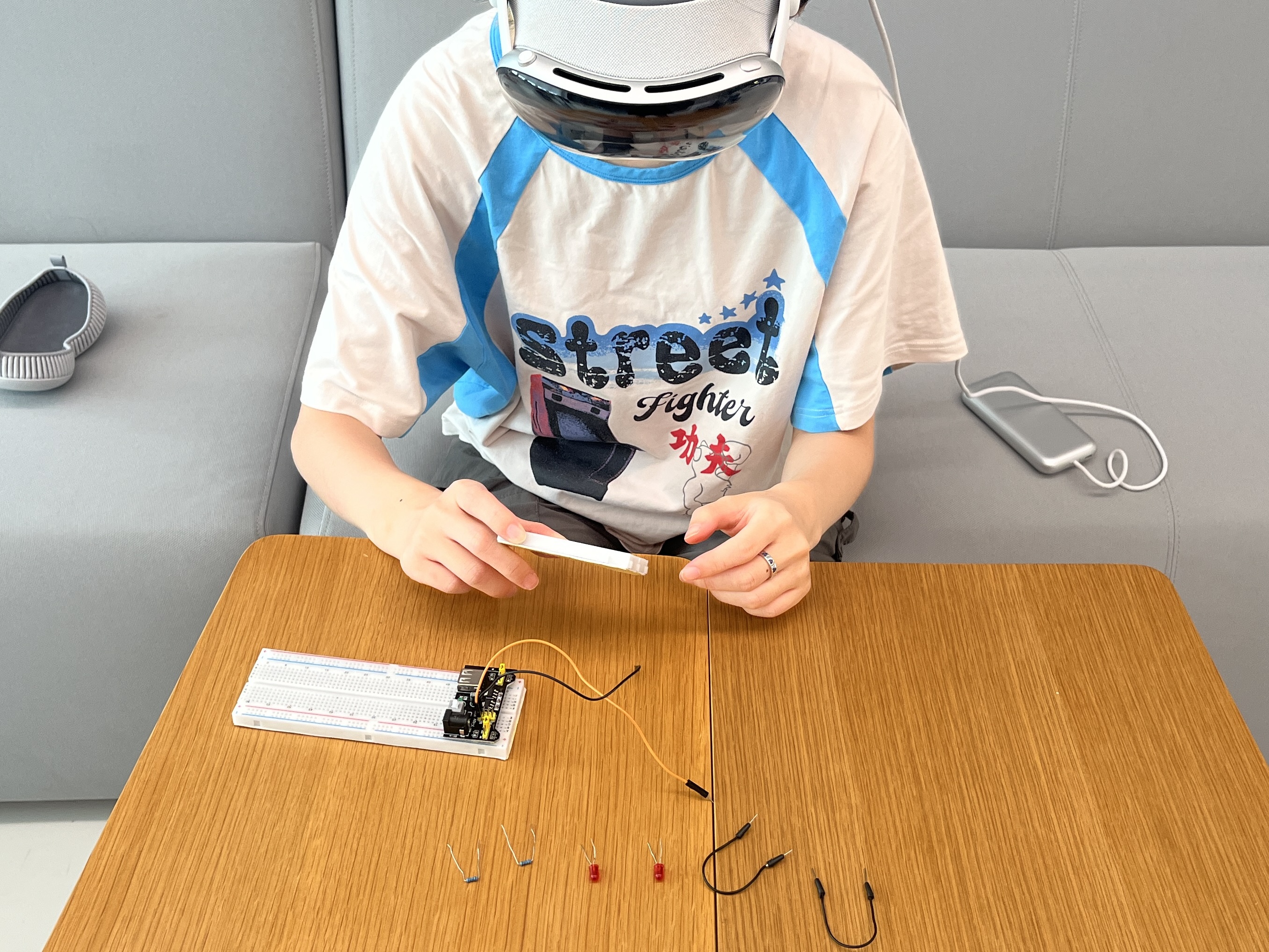}
        \label{fig:circuit}
    }
    \caption{As shown in Image (a), participants performed Task A wearing an Apple Vision Pro to follow instructions for making pour-over coffee. Equipment includes a kettle, coffee dripper, grinder, filters, and scale. As shown in Image (b), participants performed Task B wearing an Apple Vision Pro while assembling an electronic circuit using components and a breadboard.}
    \label{fig:task}
\end{figure}

The co-design session employed two distinct tasks: making pour-over coffee and assembling circuits on a breadboard (shown in Fig. \ref{fig:task}). 

\subsubsection*{\textbf{Task A: Making Pour-Over Coffee}} The pour-over coffee task exemplifies a common daily activity requiring precise step sequencing~\cite{wu2024artist}, such as grinding beans, preparing the filter, heating water, and pouring. It demands motor skills to control water flow and ensure even distribution. \remove{While the process offers immediate sensory feedback (visual and olfactory), the final outcome---taste and quality---only becomes apparent later, adding uncertainty and room for reflection. }Instructions for this task were adapted from wikiHow’s ``How to Make Pour Over Coffee'' guide\footnote{https://www.wikihow.com/Make-Pour-Over-Coffee}\remove{, adjusted to use 10g of coffee beans to reduce grinding effort.}\revision{, comprising 12 text-based steps without visual aids. The instructions were adjusted to use 10g of coffee beans to reduce grinding effort.}

\subsubsection*{\textbf{Task B: Assembling Circuits on a Breadboard}} In contrast, the circuit assembly task is more technical, commonly seen in educational or professional electronics contexts. While the order of component placement is flexible, the final configuration is critical. \remove{The feedback is immediate and binary: either both LEDs light up, or they don’t. }Instructions were based on a YouTube tutorial\footnote{https://www.youtube.com/watch?v=Bhv-Tk7l1zI}\remove{, adapted for this study.}\revision{, adapted into 8 textual steps with 4 accompanying visual diagrams showing breadboard connections and component placement, which can be challenging to explain with texts based on our formative study feedback.}

The instructions for both tasks were iteratively refined to ensure clarity and effectiveness in our setting. The tasks were chosen to develop reflective prompts applicable to various instructional contexts, as they differ in key aspects: order sensitivity, everyday versus technical focus, motor versus concept-based skills, and the presence or absence of a clear binary outcome.

\subsubsection{Method}
The co-design session was conducted in two phases for each task:

Before starting, participants were instructed to approach the tasks as they would any new, everyday task (e.g., following online instructions to cook a dish). \revision{We did not disclose that learning or understanding of the tasks was being measured during the study, as the awareness of demand characteristics can alter natural behavior and task engagement \cite{orne2017social}.
However, we disclosed the study's purpose to participants during the interviews afterward to keep them fully informed.} We emphasized that there was no single ``correct'' way to complete the tasks, encouraging natural interactions and revealing gaps between instruction-following and understanding the underlying principles.

Each participant completed both tasks twice---once without reflective prompts and once with them. \revision{To control for order effects, we counterbalanced task assignment: five participants started with Task A, four with Task B. For each task, participants first completed it without prompts, during which they identified gaps in their task completion and understanding. They then reflected on how additional support could address these gaps. After this reflection phase, participants completed the same task again with reflective prompts.} While performing the tasks, they provided feedback on the prompts and made suggestions for improvement. Participants were also asked to think aloud, offering real-time insights into their thought processes and decision-making. After every two sessions, we refined the prompts based on participant input.

After completing both tasks, each participant engaged in an individual brainstorming session with the experimenter to evaluate the prompts and generate new ideas. They reflected on moments of confusion or inefficiency and discussed how reflective prompts could have enhanced their understanding or performance. Semi-structured interviews were conducted to gather feedback on the AR system, preferences for prompts, and the impact on understanding and task performance.

Each session lasted 2 to 4 hours, with participants wearing the headset for no more than 30 minutes at a time. Given the exploratory nature of the study, no quantitative data was collected. Instead, we focused on qualitative insights from think-aloud protocols, co-design sessions, and interviews to inform the integration of reflective prompts in AR instructional systems.

\subsubsection{Results} 
The co-design sessions revealed insights about users' preferences and behaviors regarding each type of reflective prompt in AR-guided tasks.

\subsubsection*{\textbf{(R1) Critical Reflections}} \textit{Challenging Assumptions} were generally appreciated. Many (F1-7, F9) noted they often had similar questions like, \textit{``Do I have to use water of this temperature?''} or \textit{``How precise do I need to be with component placement on the breadboard?''} but \remove{didn’t pursue answers due to task focus}\revision{hesitated to pause the task to find answer}. When prompted by the system, they recognized the importance of these questions and felt motivated to seek answers as their immediate next step might be affected.

On the other hand, prompts related to \textit{Logical Reasoning} were met with mixed reactions. Many participants (F1, F2, F4, F8) expressed feelings of fatigue when encountering these prompts, as it wasn’t always clear how the logical connections between steps related to their current actions. This disconnect made the prompts feel less relevant and, at times, overwhelming.

\subsubsection*{\textbf{(R2) Reflective Pauses}} Both the \textit{Active Pause} and \textit{Opportunistic Pause} prompts were perceived as disruptive by the majority of participants (F2, F3, F4, F6-9). Prompts such as, \textit{``Let’s take a moment to reflect on...''}, were viewed negatively, even though they are commonly used in educational contexts. Participants found active breaks intrusive, interrupting their task flow. Even during natural pauses---such as waiting for water to boil---participants preferred to decide how to use the time themselves, or else they felt pressured (F2). As F7 noted, \textit{``It feels a bit bossy in an instructional system. I’d rather pause to think on my own, not because the system asked me to.''}

\subsubsection*{\textbf{(R3) Metacognitive Awareness}} Participants often ignored both \textit{Self-Assessment} and \textit{Milestone Review} prompts (F1, F3, F5-8) because the purpose of these prompts was unclear. As F1 noted, \textit{“Why do I have to think about it? It’s not like it’s evaluating me or wants an answer.”} Prompts encouraging users to revisit their reasoning, such as \textit{“What was your reasoning for placing the resistor here?”}, were typically disregarded. Most participants preferred to focus on overall task progress rather than revisiting steps they believed were correctly executed (F2, F4, F7, F8). Some found these reflective prompts time-consuming (F2, F3, F4), while others felt they disrupted their workflow (F7, F9). As F2 remarked, \textit{“It’s confusing when the system asks about a step I didn’t get wrong.”}

\subsubsection*{\textbf{(R4) Perspective Shifting}} \textit{Hypothetical Scenarios} were generally well-received by most participants (F1-6, F9). They appreciated prompts such as, \textit{“What would happen if you used more LEDs?”} or \textit{“What if we used light roast coffee beans instead of dark roast?”}, as these helped them realize the ``variables'' in the task. By toggling these ``variables'' and seeing how they are related, they built understanding and transferable knowledge that potentially improved future performance (F1-6). However, participants who viewed the task as a one-time activity were less engaged with these scenarios unless they had an immediate impact on the task (F7, F8).

In contrast, Context Transfer was largely seen as unnecessary by many participants (F3, F4, F6-9). As F4 noted, “If someone is new to the task, it’s already overwhelming to just understand the current task, let alone think about applying it to other situations.”

\begin{figure*}
    \centering
    \includegraphics[width=0.9\linewidth]{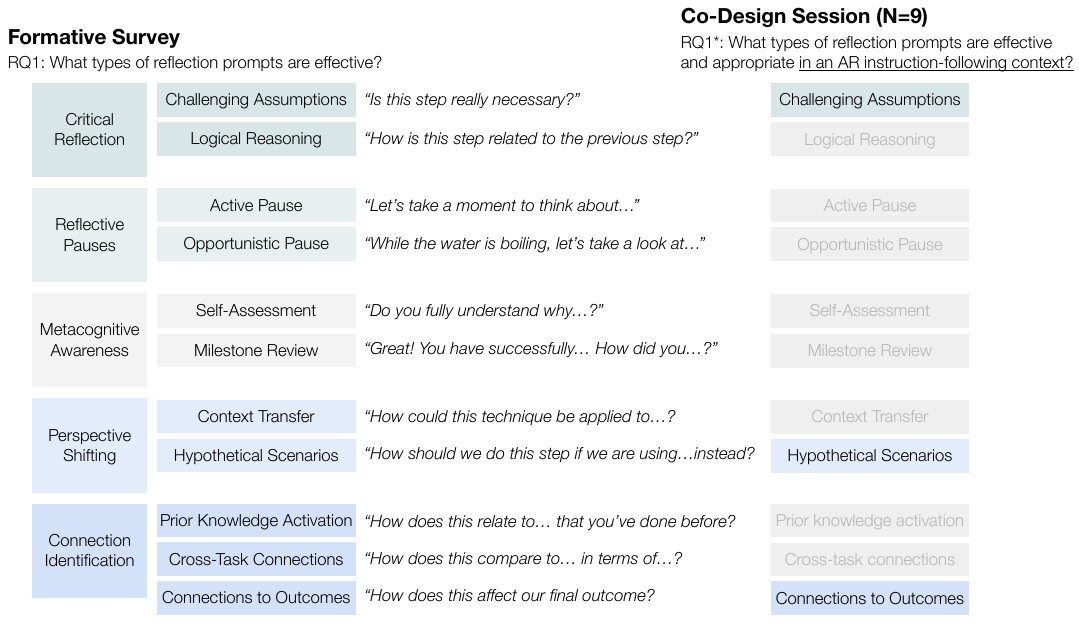}
    \caption{This figure summarizes reflection prompt types identified from the literature and evaluated for AR instruction following contexts. The left side presents five main categories of reflective prompts: Critical Reflection, Reflective Pauses, Metacognitive Awareness, Perspective Shifting, and Connection Identification, each with subtypes and example prompts. The right side shows results from co-design sessions (N=9), with colored boxes indicating the three subtypes deemed most effective and appropriate for AR instruction: Challenging Assumptions, Hypothetical Scenarios, and Connections to Outcomes.}
    \label{fig:formative}
\end{figure*}

\subsubsection*{\textbf{(R5) Connection Identification}} Most participants (F2-8) found \textit{Prior Knowledge Activation} prompts ineffective due to a lack of personalization. For instance, when the prompt matched the participant's personal experience, F1, a cooking enthusiast, resonated with a prompt like, \textit{``Controlling coffee temperature is similar to controlling oil temperature in cooking techniques,''} as he immediately connected it to frying tempura. In this study, we did not delve deeper into personalization due to the limited information available on participants' prior knowledge.

When it came to \textit{Cross-Task Connections}—linking the current skill to other potential contexts—most participants were disinterested in linking the current task to other contexts, as they were focused on immediate task completion (F2-8).

Participants were more engaged with \textit{Connection to Outcome} prompts which tied the current steps directly to the task's final outcome (F1-9). As F8 explained, \textit{``I care if this step affects the final result. If not, I could skip it.''} This suggests that framing cross-task connections in terms of immediate outcomes increases relevance and perceived value.

\subsection{Design Considerations and Updated Reflective Prompts}
\subsubsection{Design Considerations} Our study revealed several key design considerations for implementing reflective prompts in AR-guided tasks. 

\subsubsection*{\textbf{[DC1] Relate Prompts to the Current Step.}} Relevance and timing emerged as critical factors, with participants emphasizing that reflections should be closely tied to the purpose of current steps or the user's immediate situation (F1-7).

\subsubsection*{\textbf{[DC2] \remove{Avoid Burdening Users.}\revision{Optimize Timing to Reduce Cognitive Overload}}}
The system should avoid presenting reflections when the user's cognitive load is high, such as during complex task steps (F7, F8). 

\subsubsection*{\textbf{[DC3] Use Friendly, Conversational Tone.}}
Participants recommended that the tone of instructions and reflective prompts should both be conversational and friendly, respecting users' agency (F6, F7, F8). One interesting suggestion was to give reflections the persona of a friend experiencing the task alongside the user (F6).

\subsubsection*{\textbf{[DC4] Ask Brief Questions.}}
In terms of format and delivery, brief questions were perceived as more effective and desirable than statements or longer questions (F1, F2, F4, F5, F6, F9). 

\subsubsection{Updated Reflective Prompts}
Based on the insights we collected, we extracted R1, R4, and R5 as useful reflective prompts and further adapted them to the AR instruction-following context (as shown in Fig. \ref{fig:formative} and Fig. \ref{fig:prompts}):

\begin{figure*}[h]
\centering
\includegraphics[width=\linewidth]{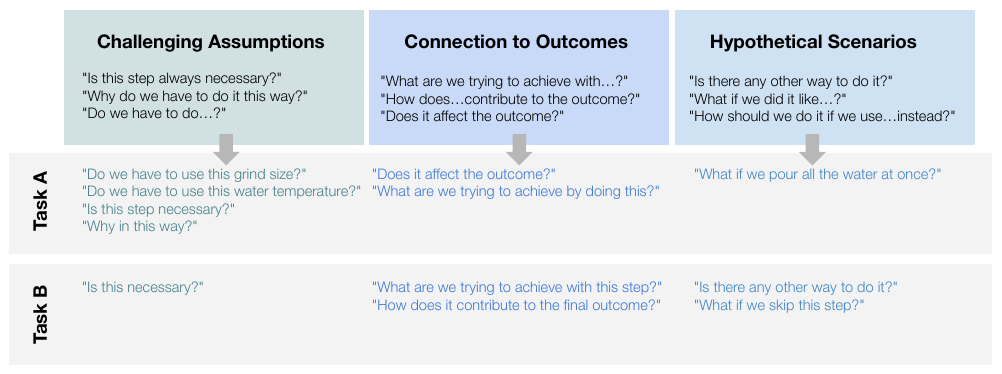}
\caption{The three types of reflective prompts we concluded with the co-design session, with example prompts and the prompts we integrated into the system we used in the evaluation study for Task A and Task B respectively.}
\label{fig:prompts}
\end{figure*}

\subsubsection*{\textbf{[R1*] Challenging Assumptions}} This type of reflection prompt encourages users to question the necessity and method of each step in the process, fostering critical thinking and deeper engagement with the task. By challenging assumptions, participants reflect on whether a given procedure is essential or if there are alternative ways to approach it. This prompt helps users become more mindful of their actions and prevents them from following instructions blindly, encouraging them to think critically about why each step is necessary.

\subsubsection*{\textbf{[R2*] Connection to Outcomes}} These prompts aim to connect specific actions to their potential outcomes, helping participants understand how each step contributes to the overall goal. This type of reflection encourages users to think about the purpose behind each action, enhancing their comprehension and decision-making process.

\subsubsection*{\textbf{[R3*] Hypothetical Scenarios}} These prompts encourage users to explore alternative ways to execute tasks, ways to execute the tasks under a different setting, or envision the consequences of different approaches. This reflective practice helps in fostering flexibility and creativity in problem-solving, particularly when users anticipate performing similar tasks in the future.

%% file: 4-evaluation-study.tex
\section{Evaluation Study}

We conducted a study (N=16) to evaluate the impact of identified reflective prompts on AR-guided task completion and the process of following instructions in AR. Participants performed the same two tasks: making pour-over coffee and assembling circuits on a breadboard, using the Apple Vision Pro.

\subsection{Participants}
Sixteen participants (\revision{P1-16, }5 female, 11 male; age: M = 24.2, SD = 3 years) were recruited from a local university. All were novices in making pour-over coffee and breadboard assembly. They received 105 CNY (15 USD) as compensation. Fifteen out of sixteen participants had prior AR experience.

\begin{figure*}[h]
\centering
\includegraphics[width=\linewidth]{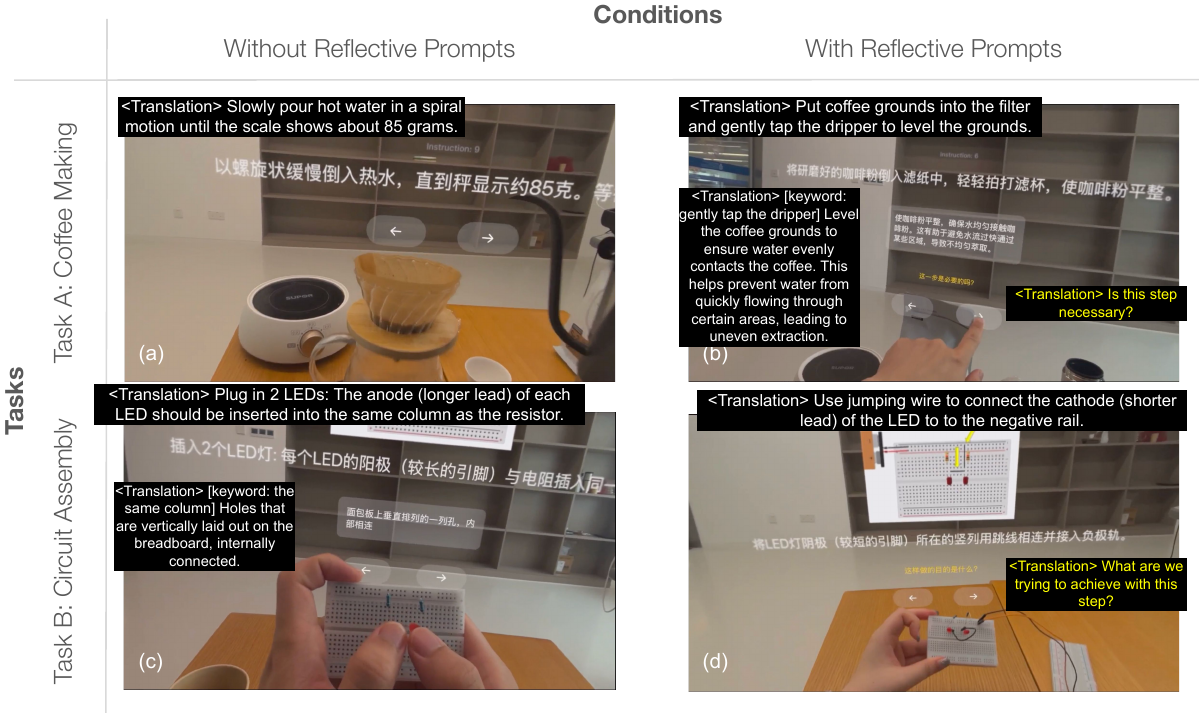}
\caption{Image (a), (b), (c), and (d) are screenshots of the participants' view in the Apple Vision Pro. Participants can click on words in the instructions to get more information wrapped in a rectangular window, as shown in (b) and (c). Interactable keywords are not distinguishable from regular instruction words. They can click on a ``Previous'' and a ``Next'' button to navigate through the instruction steps freely, as shown in (b). Reflective prompts are presented as non-interactable smaller yellow text, as shown in (b) and (d). English translations was added to the screenshots for clarity and are not part of the original interface.}
\label{fig:screenshots}
\end{figure*}

\subsection{AR Instructional System}
We iterated on the AR instructional system used in the formative study (as shown in Fig \ref{fig:screenshots}). Instructions were displayed as white text overlays, with pre-defined clickable keywords providing additional information when clicked. \revision{The keywords are clickable segments of the instructions themselves, ranging from single words to complete phrases, that users could click to get clarification or more details. For example, in the instruction ``Grind coffee beans to a medium-fine consistency (similar to granulated sugar).'', ``Grind coffee beans'' and ``medium-fine consistency (similar to granulated sugar).'' are both clickable segments, which provides more detailed information upon clicked. Task A contained 30 such clickable segments, while Task B had 22.} 
Compared to the WoZ approach applied in the formative study, this design standardized the amount of information participants could access. 
We also added ``Next'' and ``Previous'' buttons that allowed participants to navigate at their own pace. The experimenter had no control over the system during the experiment. 

In the reflective condition, yellow reflective prompts appeared in smaller font below the main instructions, 3 seconds after each new instruction step. Seven of the thirteen coffee-making steps and five of the eight circuit assembly steps included reflective prompts. These prompts were questions based on the current step, derived from the three types of prompts summarized from the formative study: \textbf{Challenging Assumptions}, \textbf{Connection to Outcomes}, and \textbf{Hypothetical Scenarios} (as shown in Fig. \ref{fig:prompts}).

\subsection{Design}
We investigated the inclusion of reflective prompts as the independent variable with two levels: with vs. without reflective prompts. 
In the reflective condition, prompts were embedded within the AR instructions to encourage deeper reflection, while the non-reflective condition provided standard AR instructions without any reflective elements.

We added task type as an additional random variable to obtain results applicable to different tasks.
We applied the same two tasks as formative study (Task A: making pour-over coffee, Task B: assembling circuits on a breadboard).
Success for Task A was defined as maintaining the correct coffee-water ratio and applying the proper water-pouring technique. For task B, success was determined by lighting up two red LEDs.

Each participant experienced each task once, one in the reflective condition and the other in the non-reflective. We counterbalanced the task and reflection condition combination: half of the participants performed the coffee-making task in the reflective condition and the circuit assembly task in the non-reflective condition, while the other half experienced the reverse combinations. Task order was also counterbalanced among participants.

We aimed to investigate the impact of reflective prompts on the overall user experience, knowledge intake of users, and their willingness to acquire relevant information.
We recorded the number of clicks on the keywords as indicators of participants' willingness to acquire additional task-related information.
We used a post-study questionnaire including four questions from the NASA Task Load Index (NASA-TLX) to measure cognitive load, and three questions from the System Usability Scale (SUS) to evaluate the AR system's usability.
For each task, we designed a multiple-choice quiz to assess how well participants remember the task procedure and understand the rationale behind the actions.  \revision{Each quiz covered three types of questions: basic conceptual understanding, factual memory recall, and knowledge transfer to new contexts. Task A's quiz contained 10 questions in total, and Task B's contained 11 questions.} 

Although participants had no prior experience with the two tasks, they might have relevant knowledge of specific quiz questions.
Therefore, we also asked participants to mark whether they knew the answer to each question before the study.
In data analysis, we excluded those questions that participants marked as known before. \revision{Across all participants, we excluded an average of 1.1 questions per participant per quiz due to prior knowledge.}
We also recorded the task outcome of each task as either success or failure. 
Through the study, we encouraged the participants to adopt the think-aloud protocol, and we also conducted an exit interview to collect their feedback. We did not measure participants' task completion time as think-aloud protocol is adopted.

\subsection{Procedure}
Upon arrival, participants were briefed on the study's purpose and provided informed consent. They then completed a pre-task questionnaire to collect demographic data and assess their pre-task interest in each task.
Participants engaged in a warm-up session to familiarize themselves with the AR device and the environment.\remove{They were contextualized to view the AR instructions as online tutorials.} \revision{Similar to the formative study, participants were instructed to approach the tasks as they would any new, everyday task (e.g., following online instructions to cook a dish). We did not provide any specific learning goals or objectives and emphasized that there was no single ``correct'' way to complete the tasks, encouraging natural interactions and revealing gaps between instruction-following and understanding the underlying principles.}

Then participants completed both tasks consecutively with a three-minute break in between. After completing both tasks, participants completed the post-task questionnaire and quizzes. We then conducted a semi-structured interview with participants to collect insights on their overall experience, encountered challenges, suggestions, and feedback for the system. \revision{During the interviews, we disclosed the study's purpose to participants to keep them fully informed.}

\subsection{Quantitative Results}

\begin{figure}[t]
    \centering
    \subfloat[Quiz Scores]{
        \includegraphics[width=0.5\columnwidth]{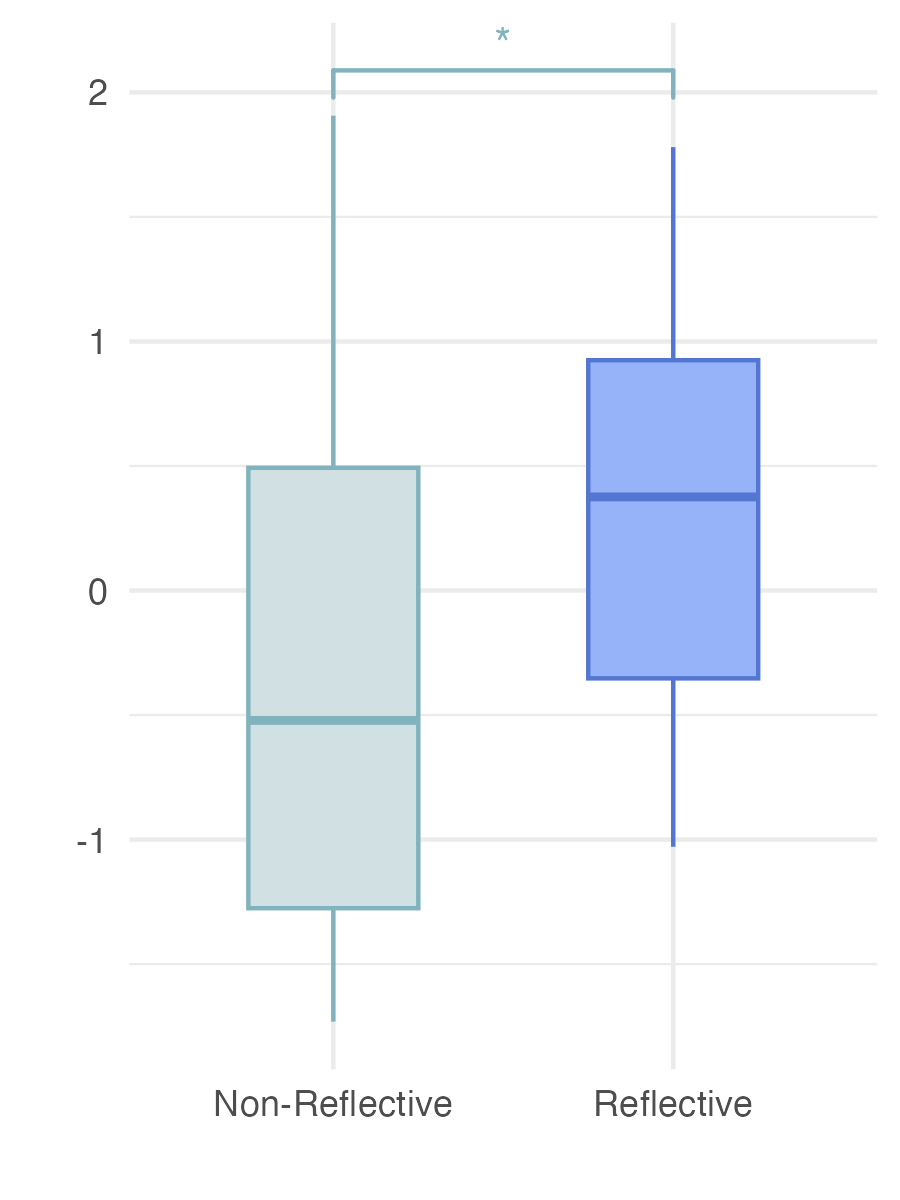}
        \label{fig:quizscore}
    }
    \subfloat[Keyword Interaction]{
        \includegraphics[width=0.5\columnwidth]{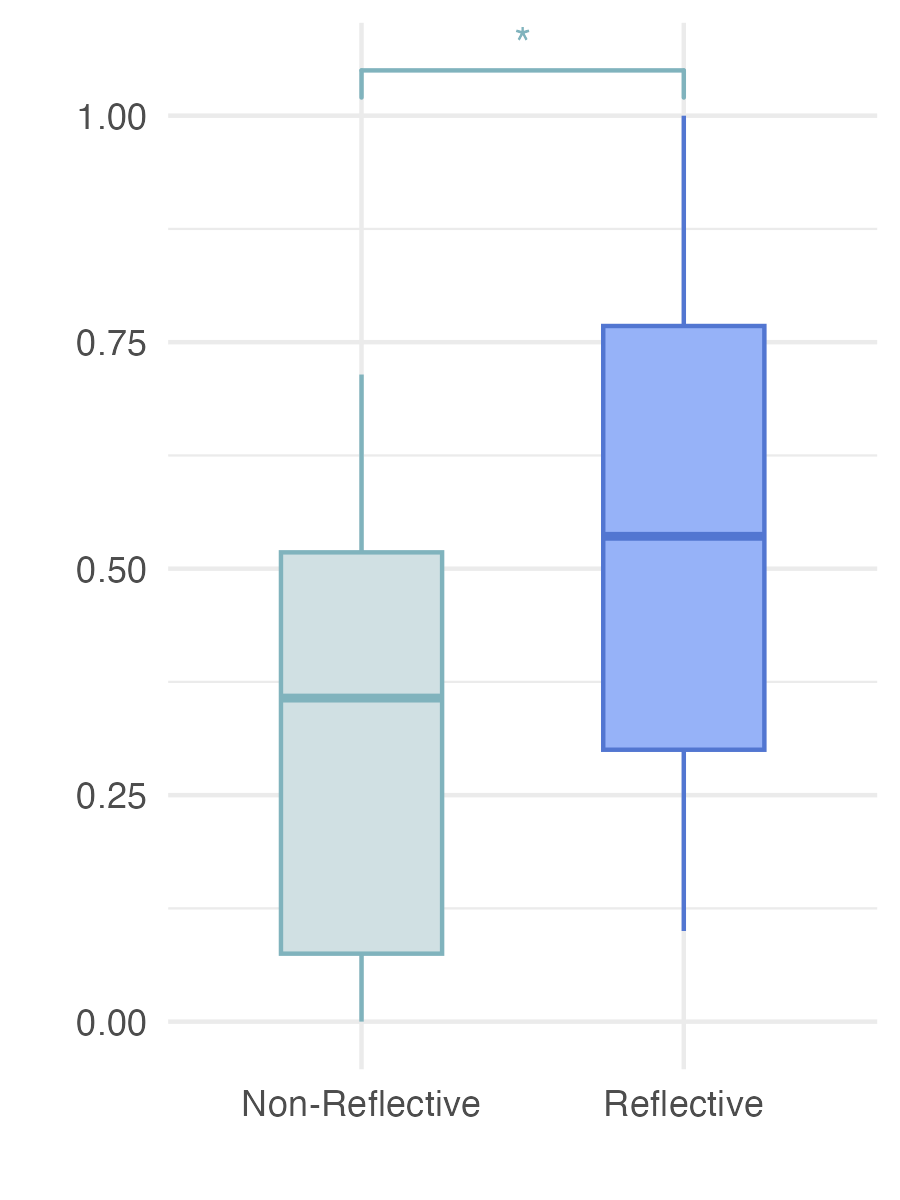}
        \label{fig:clickrate}
    }
    \caption{(a) Quiz score distribution comparing non-reflective and reflective conditions. (b) Click rate (keyword interaction rate) distribution comparing non-reflective and reflective conditions.}
\end{figure}

\subsubsection*{\textbf{Quiz Scores as a Measure of Objective Understanding}}

Quiz scores were calculated by excluding any questions where participants indicated that they already knew the answer before the task. As \textit{task type} is a random variable, we normalized quiz scores within each task using z-score normalization to account for potential differences in quiz difficulty. 
Shapiro-Wilk tests confirmed that quiz scores for each task followed a normal distribution before normalization.

A paired sample t-test showed that tasks completed with reflective prompts resulted in significantly higher objective understanding compared to those without (t(15)=2.33, p<.05, Cohen's d=0.582), with reflective prompts leading to a 0.66 standard deviation increase in quiz scores (as shown in Fig \ref{fig:quizscore}).

\subsubsection*{\textbf{Keyword Interaction as a Measure of Information Seeking Frequency}}
We calculated the total number of unique keywords clicked by participants during each task. For keywords clicked multiple times, only the first click was counted. The total click rate per task was determined by dividing the number of clicks in steps with reflective prompts by the total number of clickable keywords in those steps.

We used Min-Max normalization to account for potential differences between tasks. Shapiro-Wilk tests confirmed that click rates followed a normal distribution before normalization. A paired sample t-test revealed that click rates in reflective conditions were significantly higher than in non-reflective conditions (reflective: M=0.538, SD=0.324; non-reflective: M=0.32, SD=0.264; t(15)=2.943, p<.05, Cohen's d=0.736), indicating that participants sought task-related information 68.75\% more often when reflective prompts were present (as shown in Fig \ref{fig:clickrate}).

\begin{figure}
    \centering
    \includegraphics[width=\linewidth]{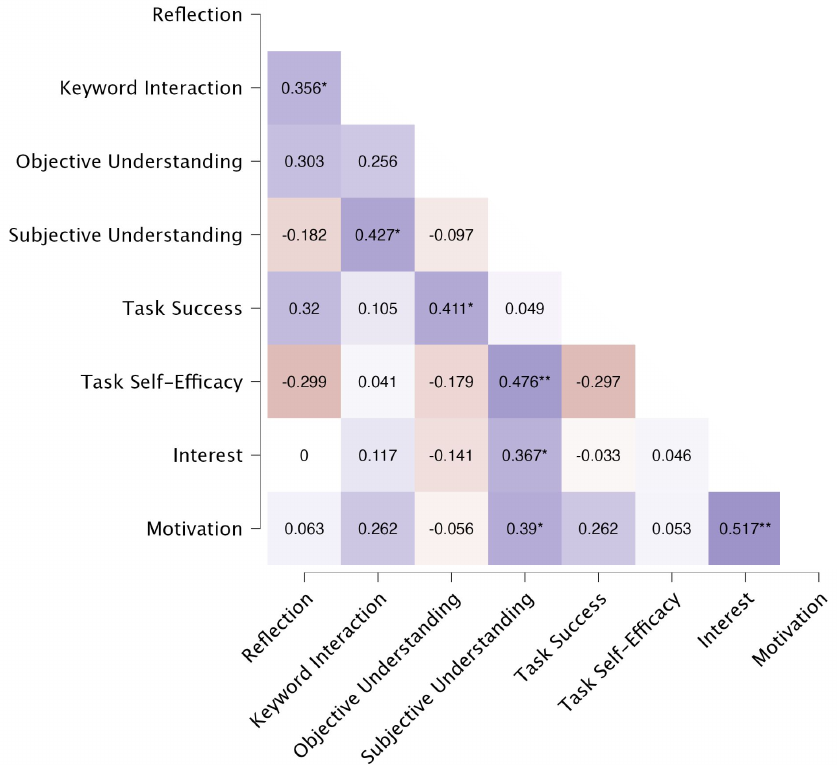}
    \caption{Correlation map showing relationships between key variables: reflection condition, keyword interaction, objective understanding, subjective understanding, task success, task self-efficacy, interest in the task and motivation to understand the underlying principles behind steps (weak correlation: |r| = 0.10 to 0.29; moderate correlation: |r| = 0.30 to 0.49; strong correlation: |r| = 0.50 to 1.00). Asterisks indicate statistical significance (* p<.05, ** p<.01).}
    \label{fig:corrmap}
\end{figure}

\subsubsection*{\textbf{Subjective Understanding}}
In the post-task questionnaire, participants rated their understanding of each task on a scale from 1 (no understanding) to 5 (perfect understanding). Shapiro-Wilk tests confirmed that subjective understanding followed a normal distribution for both tasks. For Task A, there is no significant difference in subjective understanding between the two conditions. However, for Task B, the subjective understanding is significantly lower with the presence of reflective prompts compared to without (reflective: M=3, SD=1.31; non-reflective: M=4, SD=0.93; t(7)=3.035, p<.05, Cohen's d=1.073), as revealed by a paired sample t-test. It implies the presence of reflective prompts could lower users' perceived understanding of the task. 

Subjective understanding positively correlated with keyword interaction (r=0.427, p<.05), indicating that greater information seeking was linked to better perceived understanding. It also correlated with interest in the task (r=0.367, p<.05) and motivation to understand task principles (r=0.39, p<.05) (as shown in Fig \ref{fig:corrmap}). However, subjective understanding is neither correlated with objective understanding nor task success, indicating a discrepancy between actual and perceived task understanding. 

\subsubsection*{\textbf{Task Self-Efficacy}}
In the post-task questionnaire, participants rated how confident they would be of performing each task in the future independently on a scale from 1 (not confident at all) to 5 (perfectly confident). Shapiro-Wilk tests confirmed that task self-efficacy followed a normal distribution for both tasks. Paired sample t-test showed there is no significant difference between task self-efficacy under the two reflection conditions. 

Task self-efficacy is positively correlated with subjective understanding (r=0.476, p<.01). Similarly, task self-efficacy is neither correlated with objective understanding nor task success, indicating a discrepancy between actual and perceived task efficacy. 

\subsubsection*{\textbf{Task Success}}
Success rates were higher under the reflective condition (94.8\%) compared to the non-reflective condition (68.8\%). A chi-square test of independence was performed to examine the relation between reflection conditions and task success. The relation between these variables was not statistically significant($\chi^2$(1, N=16)=2.347, p=0.126, $\Phi$=0.383). However, a moderate effect size was observed, which means the absence of statistical significance could be due to the limited sample size. Besides, task success is also positively correlated with participants' objective understanding (r=0.411, p<.05, as shown in Fig. \ref{fig:corrmap}), which implies that deepening task understanding is intrinsically aligned with the goal of task completion.

\subsubsection*{\textbf{Influence of Reflective Prompts}}
\begin{figure*}
    \centering
    \includegraphics[width=0.95\linewidth]{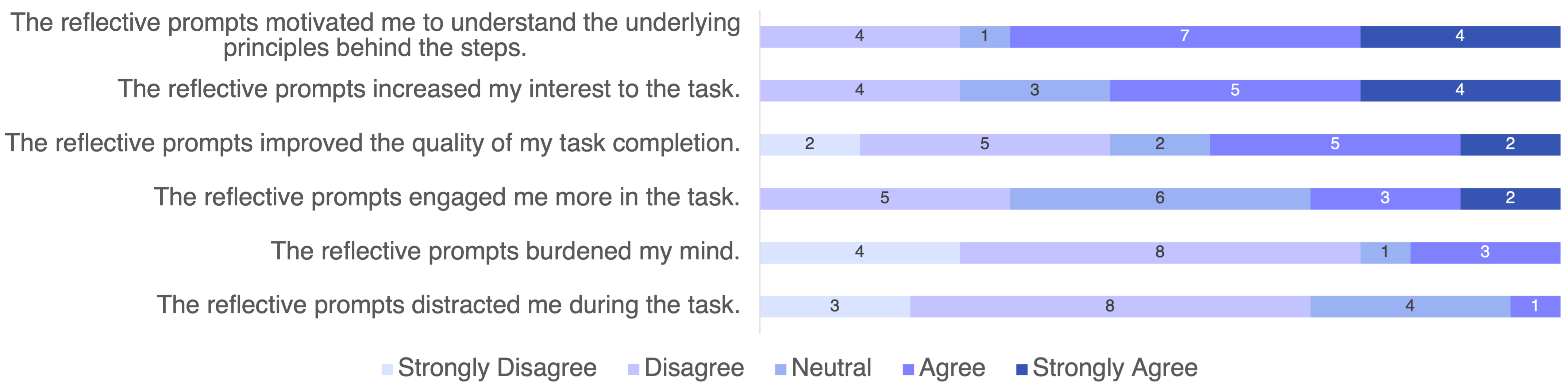}
    \caption{Opinions on reflective prompts. This Likert scale chart shows participants' responses to six statements about the impact of reflective prompts in AR instructions. Responses range from ``Strongly Disagree'' to ``Strongly Agree'' across various aspects that reflective prompts might impact, including motivation to understand principles, interest in the task, quality of task completion, engagement, mental burden, and distraction.}
    \label{fig:questionnaire}
\end{figure*}

In the post-task questionnaires, we asked the participants on a 5-point Likert scale how they perceived the reflective prompts in the tasks. Most participants either agree or strongly agree that the reflective prompts motivated them to understand the underlying principles behind the steps (11/16), and more than half of the participants either agree or strongly agree that the reflective prompts increased their interest in the tasks (9/16). Almost half of the participants either agree or strongly agree that the reflective prompts improved the quality of their task completion (7/16). Participants were overall neutral about whether reflective prompts have engaged them more in the tasks (M=3.125, SD=1.025). Most participants either disagree or strongly disagree that the reflective prompts burdened their minds (12/16) and that the reflective prompts distracted them during the tasks (11/16) (as shown in Fig \ref{fig:questionnaire}).

\subsubsection*{\textbf{Cognitive Load}}
We did not observe any statistically significant difference between the mental demand, physical demand, perceived performance, and frustration of the reflective and non-reflective conditions (as shown in Fig \ref{fig:nasasus}).

\subsubsection*{\textbf{System Usability}}
\begin{figure}[h!]
    \centering
    \includegraphics[width=\linewidth]{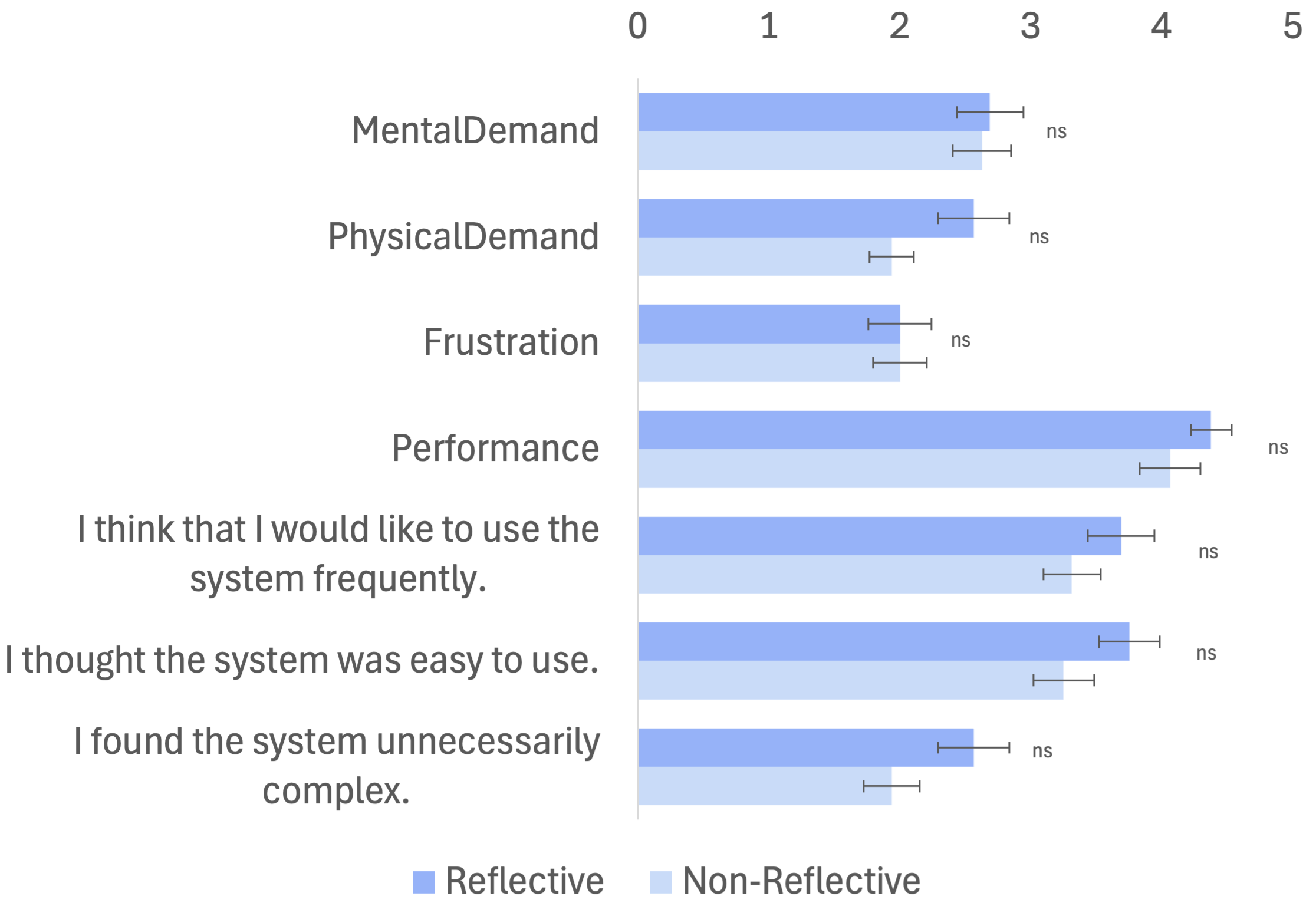}
    \caption{This horizontal bar chart displays mean ratings on a 5-point scale for four NASA TLX dimensions (Mental Demand, Physical Demand, Frustration, and Performance) and three SUS questions related to system use frequency, ease of use, and perceived complexity across the reflective and non-reflective conditions.}
    \label{fig:nasasus}
    \vspace{-10pt} 
\end{figure}

We did not observe any statistically significant difference between the system usability of the two conditions. It means the integration of reflective prompts into such an AR instructional system does not lower the system usability (as shown in Fig \ref{fig:nasasus}).

\revision{\subsubsection*{\textbf{Trust in AR Instructions}}
Participants demonstrated high trust in the AR system, with trust scores averaging 4.94 (SD = 0.25) for Task A and 4.5 (SD = 1.03) for Task B on a 5-point Likert scale. In the follow-up interviews, participants expressed strong confidence in the instructions, with comments like \textit{``I trusted the instructions 100\%''} (P2, P3, P8, P9, P11, P13) and \textit{``I never doubted it before I failed the task.''} (P2, P4)}

\subsection{Qualitative Results}
\revision{We collected qualitative data through multiple sources: video and audio recordings of participant sessions, interview transcripts, think-aloud protocols, and researcher observation notes. One author went through the session recordings, transcripts, and notes and captured qualitative insights with open thematic analysis \cite{boyatzis1998transforming}. The rest of this section presents the key themes that emerged from the analysis.}

\subsubsection*{\textbf{Reflective Prompts are Non-Intrusive to Task Completion.}}
Participants generally responded positively to the reflective prompts (15/16), appreciating their concise and unobtrusive nature. P5 remarked, \textit{``The prompts were short and not redundant.''} Others (P2, P6, P8, P14, P16) noted that the prompts were not mentally taxing and could be easily ignored if they wanted to focus on task completion.

Several participants (P6, P12) reported that the prompts didn’t significantly slow task completion, especially when they appeared during less cognitively demanding steps. P6 emphasized the importance of syncing prompts with the task’s ``rhythm'' to avoid disrupting more challenging moments. P1, P12, and P13 added that the deeper understanding gained was worth the slight slowdown, particularly during the first attempt.

\subsubsection*{\textbf{Reflective Prompts Enhance Task Execution and Understanding}} 
While participants appreciated the non-intrusive nature of the reflective prompts, they also highlighted their practical benefits. P2, P3, P4, P6, and P15-16 mentioned that the prompts encouraged them to \textit{``think before they leap''}, resulting in more deliberate, mindful task execution and improved task quality.

Some participants (P5, P11, P13) mentioned that the prompts would have motivated them to seek more information and explore further, if not limited by our prototype system. Even for intrinsically motivated users, participants felt that reflective prompts had made exploration more efficient and focused (P3, P5, P11). Additionally, some suggested that the prompts helped reinforce understanding and accelerate learning (P10, P12).

\subsubsection*{\textbf{Reflective Prompts Break the Automatic Execution of Steps.}}
Some participants felt the prompts aligned well with their natural thought processes. P2, P5, P12, and P13 noted that the prompts often mirrored their internal reflections, but appreciated seeing them explicitly as they validated fleeting thoughts or doubts that might otherwise be overlooked (P6, P10, P13). Additionally, participants like P13-15 valued how the prompts occasionally introduced new perspectives, enhancing their understanding of the task.

Most participants also observed that clear instructions without reflective prompts often led to automatic execution of steps without deeper thoughts (13/16). For example, P4 mentioned that while he initially had questions on how to connect components in series on a breadboard, he quickly forgot these questions once he proceeded smoothly with the instructions. In the interview, P4 remarked, \textit{``If there were reflective prompts in that task, I would probably have stopped and thought that through.''} Similarly, others (P2, P4, P8) suggested that reflective prompts at key moments could create a \textit{``mental pause''}, breaking the automatic flow and encouraging users to be more mindful of their actions.

\subsubsection*{\textbf{Reflective Prompts Reduce Rigid Instruction Following}} 
Although participants had the freedom to navigate the steps independently, many still adhered rigidly to the instructions, particularly in non-reflective conditions. For example, in Task A, several participants in the non-reflective condition waited for the water to boil before proceeding to the next step, even though it wasn't necessary (P7, P14). P3 followed the instructions to zero the scale without realizing the scale’s purpose was to measure the amount of water he poured.

As P8 commented on her experience of Task B in non-reflective condition, \textit{``It's only when I started to think after the task that I realized the hole alignments were not necessary.''} Reflective prompts could potentially raise users' awareness of why a step is necessary, helping them avoid rigid instruction following throughout the task completion.

\subsubsection*{\textbf{Reflection Triggered by Mistakes and Deviations}} 
In addition to reflection triggered by the system's reflective prompts, participants often initiated reflection when they sensed the potential for mistakes or realized they had deviated from the instructions. This observation is consistent with findings from our formative study. Reflective prompts that linked actions directly to outcomes were particularly effective, as they helped participants understand how certain steps contributed to either success or failure (P1, P7, P8, P9). While our prototype system did not include this feature, these insights suggest that error-prone moments could serve as valuable opportunities for reflective prompts, especially with the context-aware capabilities of AR.

\subsubsection*{\textbf{The Temptation and Caution of Mindless Instruction Following}}
Although the automatic execution of steps is often unconscious, participants recognized the tempting benefit of consciously following instructions mindlessly. Even those who valued understanding acknowledged the convenience of AR instructions, particularly in reducing cognitive load and minimizing the need for memorization. P3, P8, and P14 noted that while understanding is important, AR systems help offload mental effort, allowing users to focus on other tasks. P11 offered an analogy, comparing AR-guided tasks to driving with GPS navigation: \textit{``When you can just follow turn-by-turn directions, you can spare the attention from navigation to more important things. In real life, I'd probably be watching videos, chatting with friends, or doing something else entertaining. That’s just how people are, right?''}

Despite its convenience, most participants still expressed reservations about becoming too dependent on such systems (13/16). P1 and P2, for instance, worried that constant reliance on AR instructions might reduce engagement with the task, making it feel less immersive. P1 remarked, \textit{``I don't like it when the instructions feel too close. It feels like it's pressuring me to follow them, even though I know they’re just there to assist.''}

\subsubsection*{\textbf{Necessity to Reflect During AR-Guided Task Completion}}
Participants expressed varying opinions on the importance of reflection and understanding when following AR-guided instructions. While most acknowledged the value of reflection, some believed it was not always necessary if the system provided accurate, step-by-step guidance. For example, P8 noted that even outside of AR contexts, many people in workplaces often follow instructions mechanically, without fully understanding the reasoning behind them. She suggested that AR systems could further streamline this process, increasing productivity by reducing the need for deep comprehension of tasks.

However, the majority of participants (14/16) believed that reflection and deeper understanding are often important in AR-guided task completion, whether through intrinsic motivation or external cues, such as reflective prompts. Several participants voiced concerns about the long-term effects of not developing independent skills. P6, for instance, worried: \textit{``What if it gives me the wrong instructions, or stops providing them altogether? Would everything fall apart? That feels risky.''} P3 echoed this sentiment, stressing that understanding becomes essential in tasks where safety is involved. Some participants (P1, P3, P7, P10, P12) emphasized that reflection is particularly crucial for tasks that are frequent or complex. P14 further cautioned that overreliance on AR instructions without reflection could lead to superficial knowledge: \textit{``If systems like this become more pervasive, we might know how to do a lot of things, but we won’t truly master them. That could make us more replaceable.''}

\subsubsection*{\textbf{Priority between Task Completion and Understanding}}
Participants had mixed views on whether the system should prioritize task completion or understanding. Some believed the two goals were complementary, with understanding naturally enhancing task performance (P5, P9, P13-15). Others felt the system must carefully balance these objectives (P5, P10). P11 suggested focusing on efficient task completion, with understanding as a secondary benefit, while P1 argued that the system should actively promote understanding, especially for valuable long-term skills. Some participants, particularly in educational contexts, even proposed enforcing learning, especially for children (P13, P14).

When given a choice, participants’ preferences varied. P5 and P12 expressed, \textit{``I’ll probably always choose to understand it first.''} Most participants preferred understanding for tasks that were meaningful, frequent, or important, noting that repeatedly relying on instructions for the same task can become tedious (P3, P10). Some were motivated to understand to have control over the process (P7), while others found deeper understanding more satisfying (P9, P10, P12).

They also recognized situations where task completion should take priority—such as for temporary tasks (P2, P7, P14), trivial tasks (P13), or when speed is essential (P1, P7). Participants suggested the system should adapt to these contexts or offer different modes for users to choose from (P1, P6).

Overall, participants favored a system that balances task completion with fostering understanding (16/16).

%% file: 5-design-guidelines.tex
\section{Design Implications: Reflective AR Instructional Systems}
In this section, we synthesize insights from both the formative and evaluation studies to propose design guidelines for reflective AR instructional systems that balance task completion with user understanding\remove{ (as shown in Fig. \ref{fig:dgs})}. \revision{Even we have only explored textual reflective prompts in this paper, to build a reflective AR instructional system could use a wide range of reflective elements, such as visual cues accompanied with reflective prompts, auditory reflective prompts or simply visual cues with subtle questioning implied. In this section, we discuss the broader design implication of designing AR instructional systems with such reflective elements.}

\revision{\subsection{Design Guidelines}\label{dg}}
\subsubsection{\revision{Non-Intrusive Reflective Elements}}\label{nonintrusive}
In AR instruction contexts, task completion is often prioritized, so reflective elements must be seamlessly integrated to promote understanding without compromising efficiency. The system should provide an option for users to focus solely on rapid task completion, especially for trivial or temporary tasks. This “efficient task completion mode” allows users to bypass reflective prompts and in-depth guidance when mere task completion is prioritized. Additionally, reflective elements should align with the natural flow of tasks. They should relate directly to the current step and enhance user understanding without disrupting focus. Ideally, these elements would surface only during moments of lower task intensity, allowing users to engage with them when they have the cognitive bandwidth to do so, minimizing the risk of overwhelming users.

\subsubsection{\revision{Interactive Reflection Process}}\label{app:interactive}
Reflection should be an interactive process that deepens user understanding, particularly when users encounter challenges, make errors, or rigidly follow instructions. Although the reflective prompts used in the study were designed to be non-interactive to minimize bias, several participants (P6, P7, P9, P16) expressed that interactive prompts could foster deeper exploration. The system could offer users the option to delve into concepts more thoroughly on demand. This could include links to additional resources, explanations, or interactive examples, encouraging critical thinking about the task at hand.

Additionally, providing real-time error feedback is crucial. When users make mistakes, the system should offer immediate feedback, explaining the error and its impact on the task outcome. This kind of direct feedback helps users understand the consequences of their actions and revise their mental models, promoting deeper learning and preventing repeated errors.


\subsubsection{\revision{Encouraging Necessary Understanding}}\label{app:necessaryunderstanding}
The system should promote deeper understanding in scenarios where it is most beneficial, such as during high-stakes, complex, or frequently performed tasks that are valuable to users. In these cases, encouraging users to engage with core concepts can enhance their long-term proficiency and independence. However, for routine or less critical tasks, the system should remain flexible, allowing users to prioritize efficiency and bypass reflection. Some users, particularly those less intrinsically motivated, value motivational reinforcement. For example, P1 and P4 noted the usefulness of motivational nudges, especially when the tasks were important or frequently encountered.

\subsubsection{\revision{Respecting User Autonomy}}\label{app:autonomy}
Maintaining user control over the AR instructional system is essential, particularly regarding reflective elements and the level of guidance provided. Respecting user autonomy enhances satisfaction and ensures flexibility, allowing the system to cater to individual needs and preferences. Reflective elements should be framed in a friendly, conversational tone that encourages engagement without making users feel pressured or overwhelmed. Additionally, offering customizability is key. Users should be able to tailor the content and depth of both instructions and reflective elements, ensuring that the system adapts to their preferences while giving them the freedom to shape their experience.

\subsubsection{\revision{Adaptive and Context-Aware Guidance}}\label{adaptive}
AR instructional systems that encourage user understanding should dynamically adjust their level of guidance based on user performance, task complexity, and familiarity. When users are unfamiliar with tasks or face difficulties, more support should be offered. Conversely, as they gain proficiency, assistance should gradually decrease, but not in a way that hinders task completion. Unlike traditional scaffolding in educational contexts, where reducing support is intended to challenge the learner, the goal here is to maintain ease of use while promoting independence. This adaptive approach encourages users to move beyond reliance on instructions as a ``crutch,'' helping them develop greater independence and a deeper understanding of the task. By setting the expectation that guidance will taper off over time, the system encourages greater self-sufficiency and deeper understanding without introducing unnecessary complexity or frustration.

\subsection{\revision{Reflective AR Instructions in Broader Scenarios}}
\revision{
While our empirical studies focused on two specific tasks, we explore how the concept of incorporating reflective elements into AR instructions might extend to other scenarios. To facilitate discussion and inspire future research, we propose examining tasks through the lens of potential \textbf{reflection values}---the specific benefits that reflection can bring to task performance. By identifying where reflection adds the most value, designers can adapt the guidelines in Section \ref{dg} to optimize these benefits and create more engaging, effective instructional experiences.}

\subsubsection{\revision{Reflection Values}}\label{reflectionvalues}
\revision{We identify five main aspects where reflection can enhance user learning and performance. Tasks may engage multiple reflection values depending on their complexity and context. }

\begin{figure*}
    \centering
    \includegraphics[width=1.0\linewidth]{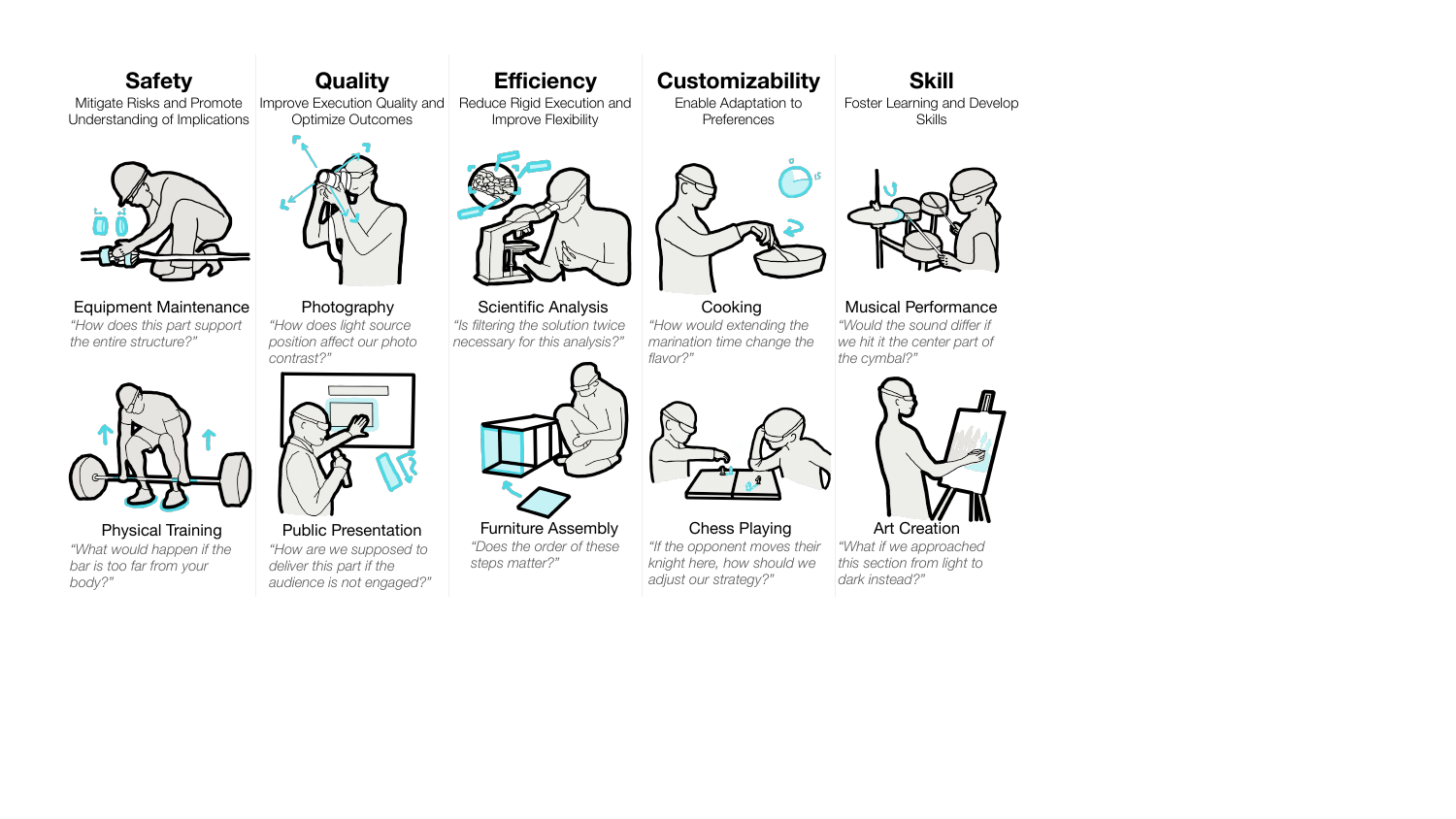}
    \caption{\revision{A range of representative reflective AR instruction scenarios organized by five core reflection values (Safety, Quality, Efficiency, Customizability, and Skill). Each scenario shows physical tasks with virtual guidance (indicated in blue), with a sample reflective prompt that encourages deeper task understanding. Tasks may emphasize different reflection values depending on contexts and personal objectives.}}
    \label{fig:enter-label}
\end{figure*}

\subsubsection*{\textbf{\revision{Safety}}}
\revision{
Reflection supports safety in two key ways. First, it mitigates the risk of users mindlessly following AR instructions, which are inherently limited by hardware constraints in detecting potential hazards (e.g., tactile feedback, odors, or specific environmental factors). Humans possess instincts and multi-sensory capabilities to assess risks, and over-reliance or overtrust in AR may erode these critical skills.

Second, reflection enhances users' understanding of the safety implications of specific steps, ensuring they grasp the rationale behind key actions and anticipate the potential consequences of improper execution. For example, in rock climbing, the PBUS (Pull, Brake, Under, Slide) belaying technique\footnote{https://www.wikihow.com/Belay} requires precise hand positioning and rope control. While AR can visually demonstrate the steps, users must reflect on \textit{why} these seemingly arbitrary hand positions are essential to maintaining safety.

When the primary reflection value for a task is safety, system designers may consider reinforcing the guideline \ref{app:necessaryunderstanding} by incorporating \textbf{non-dismissable critical reflective elements} and exploring ways to increase the \textbf{frequency and prominence} of reflection prompts. However, it is equally important to carefully evaluate the timing of these reflective elements to ensure they do not split users' attention or inadvertently introduce potential risks during operation.}
\subsubsection*{\textbf{\revision{Quality}}}
\revision{
Reflection can also improve the quality of task performance. While instructions provide a clear path to completing a task, they do not always lead to the best possible outcome. Novice users, in particular, may struggle to establish the mental frameworks necessary to evaluate task quality if they lack a deeper understanding of the task. Without this understanding, they may fail to recognize opportunities to optimize their performance beyond simply following the prescribed steps.

For example, in photography, novice users might rigidly follow parameters and perspectives given by AR instructions without understanding their purpose—such as balancing elements for visual appeal. This can prevent them from recognizing when breaking these guidelines, like centering a subject or using negative space that might better capture the mood or story of a scene. By reflecting on the principles behind these parameters, users can adapt their framing dynamically, producing photos that are not only technically correct but also more expressive and impactful.

When task performance quality is the primary reflection value, system designers could provide \textbf{real-time feedback on errors and improvement opportunities}. Following guideline \ref{app:interactive}, this feedback might highlight ways to enhance current performance and refine future efforts. Encouraging users to explore alternative methods can also help them understand how different approaches impact outcome quality, leading to better results.}

\subsubsection*{\textbf{\revision{Efficiency}}}
\revision{Reflection can improve task efficiency by helping users recognize when rigid adherence to instructions is unnecessary. While AR instructions aim for precision, this can sometimes lead to inefficiency if users follow every detail without understanding the rationale. 

For example, in our formative study, F3 attempted to add water until the scale showed 85g but overshot the target. To correct this, they removed coffee, dropping the weight below 85g, and then added water again. This back-and-forth adjustment wasted time and diluted the coffee, whereas a simple adjustment to the subsequent pour could have resolved the issue.

When efficiency is the primary reflection value, system designers could follow guidelines \ref{nonintrusive} and \ref{adaptive} by providing non-intrusive reflections that \textbf{minimize disruption} and \textbf{reduce overly detailed instructions} based on context and user expertise. Additionally, adopting guideline \ref{app:necessaryunderstanding} can help users distinguish between critical and flexible steps.}
\subsubsection*{\textbf{\revision{Customizability}}}
\revision{Reflection can enhance customizability by helping users identify opportunities to adapt instructions to their preferences. While AR systems may learn user preferences over time, they could still struggle to provide highly customized instructions in novel tasks. Without understanding the principles behind each step, users may be unable to safely or effectively customize tasks.

For instance, in cooking, understanding the purpose of specific steps allows users to make informed modifications for desired flavors or nutritional goals while preserving the dish's essential characteristics. This adaptability extends beyond the task, enabling users to transfer these skills to other contexts. Guided reflection during task execution helps users recognize which elements can be modified and how to make purposeful adjustments.

When customizability is the primary reflection value, system designers could follow guideline \ref{app:interactive} to inspire users with \textbf{customization possibilities} and guideline \ref{adaptive} to create instructions that \textbf{dynamically adapt to users’ evolving preferences}.}
\subsubsection*{\textbf{\revision{Skill}}}
\revision{Reflection supports skill development by helping users move beyond task completion to understand underlying principles and build mental models for independent problem-solving and decision-making. By reflecting, users can recognize patterns across tasks and develop a deeper understanding of cause-effect relationships.

Skill development is almost always beneficial, even for seemingly trivial tasks, like furniture assembly, as it enables faster, more independent performance in the future. However, engaging in reflective learning is ultimately a personal choice. While the benefits of skill development are clear, respecting user autonomy in choosing their level of engagement is essential.

When skill development is the primary reflection value, all guidelines can be applied to encourage \textbf{meaningful and interactive exploration with scaffolding instructions}. However, designers should prioritize guideline \ref{app:autonomy} to \textbf{respect users’ preferences} for how deeply they wish to engage in skill-building.
}

\subsubsection{\revision{Limitations of AR Instructions}}
\revision{While AR instructions can effectively support many tasks, their effectiveness could vary significantly across different contexts. Certain fundamental limitations of AR technology make some tasks less suitable for AR-based instruction or reflective elements. Understanding these limitations is crucial for determining where reflective AR instruction systems can be most effectively deployed.
Here we identify several key factors that may limit the effectiveness of AR instructions in tasks:
}
\subsubsection*{\textbf{\revision{Tacit Knowledge and Physical Intuition}}}
\revision{Tasks that heavily rely on tactile feedback and physical intuition or information from other channels such as smell and taste, may be difficult to guide through AR alone. The subtle sensations and muscle memory required in these activities cannot be fully captured or conveyed through visual overlays.}

\subsubsection*{\textbf{\revision{Time-Critical Responses}}}
\revision{In situations requiring split-second decisions or immediate physical reactions, such as sports performance or emergency responses, AR guidance may introduce unwanted latency or cognitive overhead. The time needed to process AR information could interfere with natural reflexes and intuitive responses.}

\subsubsection*{\textbf{\revision{Social Interaction}}}
\revision{While AR can provide cues for social situations, over-reliance on AR instructions in interpersonal scenarios might hinder authentic human connection and emotional understanding. Tasks like counseling or conflict resolution require nuanced reading of social cues that AR systems may not fully capture.}

%% file: 6-discussion.tex
\section{Discussion}

\subsection{\remove{High }Trust in AR Instructions} 
\remove{Participants demonstrated strong trust in the AR system, with trust scores averaging 4.94 (SD = 0.25) for Task A and 4.5 (SD = 1.03) for Task B. Many participants expressed complete confidence in the instructions, stating, ``I trusted the instructions 100\%'' and ``I never doubted it before I failed the task.'' Notably, all participants (16/16) indicated they were generally skeptical of content from large language models (LLMs) like ChatGPT, yet did not share the same caution toward AR instructions. This trust}\revision{While all participants (16/16) reported being generally skeptical of content from large language models (LLMs) like ChatGPT, they did not express similar skepticism toward the AR instructions. The high trust level} likely stems from the immersive nature of AR, where the overlays felt seamlessly integrated with the physical task. Participants described the instructions as part of the physical tasks, which likely reinforced their perception of instruction accuracy (P1, P6).

\remove{However, this high trust could also be due to the interface layout of the prototype system.}\revision{Besides, the interface design of our prototype system may have further reinforced this trust.} The simple AR interface and the lack of visible information sources may lead users to accept instructions uncritically. This may be worth exploring in future studies, as our interview findings suggest that increased trust in AR guidance often leads to greater reliance on the instructions, resulting in more fluent and confident task execution. When combined with the potential for attention tunneling effect in AR \cite{yeh2001display}, this trust and reliance could raise safety concerns, as users may become more prone to overlooking mistakes and hazards during task performance. \revision{While reasonable trust can be beneficial, novices often struggle to form appropriate skepticism toward AI instructions. Beyond the five main reflection values proposed in Section \ref{reflectionvalues}, reflective prompts could encourage users to question not only tasks and steps, but also how the system derives its instructions \cite{liao2021question}. This could help users better understand the system's capabilities and limitations, potentially enhance system explainability, and foster balanced trust between novice users and AR instructional systems.}

\subsection{The Disconnect between Actual and Perceived Task Competence}
Our evaluation study revealed a notable disconnect between participants' objective task competence (objective understanding and task success) and their subjective competence (subjective understanding and task self-efficacy). While objective understanding and task success were positively correlated (r=0.411, p<.05), and subjective understanding and task self-efficacy were similarly linked (r=0.476, p<.01), there was no significant relationship between objective and subjective measures. Interestingly, task self-efficacy was slightly, though not significantly, negatively correlated with task success (r=-0.299), and a similar negative correlation appeared between self-efficacy and the reflection condition (r=-0.299) (as shown in Fig. \ref{fig:corrmap}). 

This disconnect suggests that participants' confidence in their abilities may not align with, and could even be negatively correlated with their actual task performance\revision{, which aligns with the Dunning-Kruger effect \cite{kruger1999unskilled} where less competent individuals tend to overestimate their abilities}. \remove{Additionally, the presence of reflective prompts could potentially contribute to lower perceived competence. A possible explanation is the limited information available in the system. When participants were prompted to reflect, they sometimes developed doubts beyond the reflective questions which were not fully addressed through keyword interactions or logical reasoning. Moreover, these prompts helped them realize the underlying depth of the task, which may have contributed to their uncertainties. Given that participants only performed the task once, this initial exposure may have limited their confidence.}\revision{In our study, reflective prompts may have contributed to lower perceived competence in two ways. First, they might have triggered questions in participants that exceeded the prototype system's capacity to address through its limited keyword interactions and participants' reasoning capabilities. Second, by revealing the task's underlying depth, the prompts led participants to recognize knowledge gaps, potentially affecting their confidence. This effect may have been amplified by participants' single initial exposure to the task.}

Despite this, self-efficacy remained moderate in both the reflective (M=3.563, SD=1.209) and non-reflective (M=3.813, SD=0.911) conditions. A paired sample t-test confirmed that reflective prompts did not significantly lower self-efficacy or subjective understanding.

\remove{Future work should explore this disconnect further, particularly examining whether reflective elements could undermine self-efficacy over time and investigating the underlying mental mechanisms contributing to this effect.}\revision{Future work could investigate this competence-confidence disconnect longitudinally to determine whether reflective elements help mitigate the Dunning-Kruger effect by cultivating more accurate self and task assessment, or potentially lead to frustration with self-efficacy undermined. Understanding this relationship could inform how to balance reflection and confidence-building in instructional systems.}

\subsection{The Challenge of Learning from Instructions}
Following instructions is inherently a linear process, guiding users through a single, predefined solution path. However, achieving a deeper understanding often requires a more nuanced, hierarchical grasp of concepts. In both the formative and evaluation studies, we found that while most participants could follow instructions to connect two components on a breadboard, they often failed to realize that the same pattern should be avoided when the goal was not to connect components. As noted by F1, F2, and F7, instructions alone do not promote reverse thinking or flexible problem-solving.

Developing comprehensive task proficiency requires more than merely following instructions—it involves active observation, reflection, and experimentation based on abstract conceptualization. While reflective prompts do not directly lead to complete mastery or structured knowledge, they nudge users to step away from the linear task execution process and engage with it from a higher-level perspective. This shift in perspective can stimulate conceptual understanding and curiosity to experiment further.

Reflecting on the three types of effective prompts identified in the formative study---challenging assumptions, connecting actions to outcomes, and exploring hypothetical scenarios---we found that they provide a more effective approach to the instruction following than the traditional linear mindset of ``What is it?'' and ``How do I do it?'' These prompts instead foster a more critical thought process: ``Do I have to do it this way?'', ``How does this contribute to my desired outcome?'', and ``How might I approach this in different circumstances?''

This mindset encourages critical thinking, skepticism toward instructions, and flexibility in problem-solving, helping users avoid rigid, overly precise adherence to steps. It also aids in building correct mental models and identifying transferable knowledge, both of which are essential for skill development. Ideally, this reflective mindset would become a natural part of task execution.

Computer interfaces have the potential to shape human thinking patterns. With the rise of large language models (LLMs), many users have become increasingly reliant on technology, often at the expense of critical thinking and active understanding \cite{bastani2024generative, abbas2024harmful}. If this trend were to extend to AR instructions as AR technology becomes more pervasive, users might be reduced to mere executors of instructions, losing the ability to make independent decisions when needed. Therefore, we argue that instructional interfaces have the responsibility to influence users in the opposite direction—toward \textit{more mindful, critical, and intelligent engagement} with tasks.

\subsection{\revision{Rethinking Metacognitive Demands of Generative AI in AR Instructional Contexts}}
\revision{Recent work has examined the metacognitive demands of working with GenAI (Generative AI), primarily in screen-based interfaces where interactions resemble a manager delegating tasks to a team \cite{tankelevitch2024metacognitive}. However, AR instructional contexts present a unique case for metacognitive investigation due to their distinct interaction patterns.

In AR environments, GenAI can understand users' context through both diegetic (users' current context, predicted intentions) and non-diegetic (explicit commands) inputs. Unlike traditional GenAI interactions focused on task delegation, AR instruction-following requires users to process and execute guidance themselves actively. This shift from delegation to guided execution introduces different metacognitive demands, including self-awareness of past, present, and future actions, and appropriate adjustment of confidence and reliance based on continuous assessment of instruction and execution quality.

While increased metacognitive demands in human-GenAI interaction could be viewed as a challenge to address \cite{tankelevitch2024metacognitive}, they might actually be beneficial in AR instructional contexts---and potentially insufficient. Research shows that easily processed information triggers less cognitive processing compared to more challenging information \cite{ackerman2019heuristic}. When instructions are too easy to follow, they may reduce both users' cognitive processing during the task and beneficial metacognitive opportunities. This fundamental tension between smooth task completion and promoting deeper task understanding further justifies our use of external reflective prompts.}



\subsection{Limitation and Future Work}

\subsubsection{Evaluation of Knowledge Acquisition}
In the evaluation study, we employed quizzes as the primary method for evaluating knowledge acquisition and retention. While quizzes are effective for assessing factual recall and procedural understanding, \remove{they may not fully capture the depth of users’ cognitive engagement or their ability to apply learned knowledge in real-world contexts}\revision{their limitations in capturing deeper cognitive engagement and real-world applicability remain.}

To address this limitation, future research could incorporate \remove{a variety of assessment}\revision{alternative} methods beyond quizzes\revision{, such as performance-based assessments where participants independently complete tasks without AR assistance, offering clearer insights into their application of learned knowledge.}\remove{For instance, performance-based assessments where participants are asked to complete tasks without AR assistance could provide a clearer picture of their ability to apply learned knowledge independently.} \revision{Additionally,}\remove{Alternatively,} think-aloud protocols during follow-up tasks, where participants verbalize their thought processes, could offer deeper insights into their understanding and decision-making. Longitudinal studies that track knowledge retention over extended periods would \remove{also be beneficial to}\revision{help} assess how well participants retain and transfer knowledge to new contexts.

\subsubsection{Generalizability of the Task}
\revision{Our studies employed only two tasks, which may not fully represent the diversity of tasks encountered in AR instructional systems.}\remove{Our studies employed only two tasks: making pour-over coffee and assembling circuits on a breadboard. While these tasks allowed us to compare two distinct activities, they do not capture the full diversity of tasks users may encounter when using AR instructional systems. Both tasks were relatively structured and followed well-defined steps, which may not represent more open-ended or creative tasks where instructions are less rigid or the outcomes are more variable.} Additionally, participants may have had varying levels of relevant knowledge of these tasks, which could also influence their engagement with the reflective prompts and instructions. 

Moreover, the binary success/failure outcome used in this study does not fully capture the nuanced differences in task performance. Future studies should consider \revision{including more diverse metrics, such as errors made, corrective actions, task efficiency, and unnecessary over-precision. These measures provide deeper insights into how reflection influences task understanding and outcomes.}\remove{incorporating more detailed performance metrics, such as the number and types of errors made during the task, corrective actions taken, efficiency of task completion, or instances of unnecessary over-precision in following instructions. These measures can provide a richer understanding of how users' performance is influenced by reflection, shedding light on the specific factors that contribute to differences in the overall task understanding and outcome.}

\subsubsection{Generalizability of the Reflective Prompt Types}
The reflective prompts in this study were developed based on two specific task scenarios—coffee-making and circuit assembly. While these tasks span different domains, they may not encompass the full range of contexts where reflective prompts could be applied. \revision{Future research could consider expanding, investigating, and evaluating the reflective prompt types and reflection values in Section \ref{reflectionvalues} in more application scenarios.}\remove{Consequently, the effectiveness of these prompts may be limited to tasks with similar characteristics, such as procedural tasks or those involving physical interactions. Therefore, their generalizability to other tasks may necessitate adapting the prompts to fit varying task structures and user needs. Investigating how reflective elements can be tailored and scaled across different AR applications will be essential for fully understanding their impact and potential to promote deeper learning and comprehension in a wider array of task environments.}

\subsubsection{Limited Observations of Behaviors}
One of the significant limitations of this study is the short-term nature of our observations and limited sample size. Participants only interacted with the AR instructional system once, and we were unable to observe long-term usage patterns or the development of habitual behaviors. User interactions with AR technology may evolve over time \revision{as familiarity increases and users integrate the system into their everyday workflow.}\remove{, as users become more familiar with the system and integrate it into their workflow. While participants reported that they would not rely heavily on the system in the future, their actual behavior may differ after repeated exposure or once they encounter more complex or unstructured tasks.}

\revision{Furthermore, initial reactions to reflective prompts may not reflect long-term engagement. Prompts seen as disruptive during early use might become more valuable over time, while initially well-received prompts might lose effectiveness with repeated use.}\remove{Furthermore, participants' initial reactions to reflective prompts may not accurately represent their long-term engagement with them. Reflective prompts that seemed disruptive during the first use might become more valuable over time as users grow accustomed to them and begin to see the benefits of deeper reflection. Conversely, prompts that were initially well-received might become less effective or even ignored with repeated use.}

%% file: 7-conclusion.tex
\section{Conclusion}
This paper explored the role of reflective prompts in enhancing understanding during AR-guided task completion. Through a formative survey and co-design sessions, we identified and refined three types of prompts: \textbf{Challenging Assumptions}, \textbf{Connecting to Outcomes}, and \textbf{Hypothetical Scenarios}. Our evaluation study showed that these prompts significantly improved participants' objective task understanding and encouraged them to acquire more task-related information. However, the prompts could potentially lower perceived understanding. Drawing on both quantitative and qualitative data, we proposed design guidelines for reflective AR instructional systems and discussed how they could generalize to broader use case scenarios. This work contributes to the growing field of AR as an instructional medium, demonstrating how embedded reflective prompts can influence both user comprehension and task performance.

%% file: acknowledgements.tex
\begin{acks}
    This research was funded part by the NSERC Discovery Grant RGPIN-2021-02857 and JST PRESTO Grant Number JPMJPR23I5. We also thank all of the participants for our user study.
\end{acks}